\documentclass[fleqn]{2023SCGE}
\usepackage{hyperref}
\usepackage{ulem}
\usepackage{multirow}
\usepackage{booktabs}
\usepackage{subfigure} 
\usepackage{bm}

\begin{document}
\ensubject{subject}
\ArticleType{Article}
\SpecialTopic{SPECIAL TOPIC: }
\Year{*}
\Month{*}
\Vol{*}
\No{*}
\DOI{??}
\ArtNo{000000}
\ReceiveDate{****}
\AcceptDate{****}
\AuthorCitation{Dengke Zhou et al.}
\title{A comprehensive search for Long and Short Periodic Features from an Extremely Active Cycle of FRB 20240114A}

\author[1]{Dengke Zhou}{}
\author[2,3,4]{Pei Wang\footnote{Corresponding author. Email: wangpei@nao.cas.cn}}{}
\author[1]{Jianhua Fang}{}
\author[2,3,5,4]{Weiwei Zhu\footnote{Corresponding author. Email: zhuww@nao.cas.cn}}{}
\author[6,7,8]{Bing Zhang\footnote{Corresponding author. Email: bzhang1@hku.hk}}{}
\author[9,2,4]{Di Li\footnote{Corresponding author. Email: dili@tsinghua.edu.cn}}{}
\author[1,10]{\\Yi Feng\footnote{Corresponding author. Email: yifeng@zhejianglab.org}}{}
\author[11,12]{Yong-Feng Huang}{}
\author[13,2,14,15]{Ke-Jia Lee}{}
\author[2,5,4]{Jinlin Han}{}
\author[16,17]{Yuan-Chuan Zou}{}
\author[2,5]{\\Jun-Shuo Zhang}{}
\author[18,19]{Shuo Xiao}{}
\author[20]{Rui Luo}{}
\author[16]{Long-Xuan Zhang}{}
\author[21,3]{Tian-Cong Wang}{}
\author[2,5]{Wanjin Lu}{}
\author[2,5]{\\Jinhuang Cao}{}
\author[22]{Wenfei Yu}{}
\author[23]{Bing Li}{}
\author[24]{Chen-Chen Miao}{}
\author[25]{Jintao Xie}{}
\author[1]{Yunchuan Chen}{}
\author[1]{Han Wang}{}
\author[7,8]{\\Yuanhong Qu}{}
\author[1]{Huaxi Chen}{}
\author[2,5]{Yuhao Zhu}{}
\author[14,5]{Shuo Cao}{}
\author[2]{Xiang-Lei Chen}{}
\author[11]{Chen Du}{}
\author[21,3]{He Gao}{}
\author[14,5]{\\Yu-Xiang Huang}{}
\author[17]{Ye Li}{}
\author[26,27]{Jian Li}{}
\author[28]{Dong-Zi Li}{}
\author[21,3]{Lin Lin}{}
\author[2,5]{Xiaohui Liu}{}
\author[29,30]{Jia-Wei Luo}{}
\author[2]{\\Jiarui Niu}{}
\author[31]{Chen-Hui Niu}{}
\author[2,5]{Qingyue Qu}{}
\author[16]{Shiyan Tian}{}
\author[2,3,5,4]{Chao-Wei Tsai}{}
\author[11,12]{Fayin Wang}{}
\author[2,5]{\\Yi-Dan Wang}{}
\author[5]{Wei-Yang Wang}{}
\author[2]{Bojun Wang}{}
\author[32,33]{Suming Weng}{}
\author[11]{Qin Wu}{}
\author[2]{Zi-Wei Wu}{}
\author[2]{Heng Xu}{}
\author[2,4]{\\Aiyuan Yang}{}
\author[34]{Yuan-Pei Yang}{}
\author[32,33]{Shihan Yew}{}
\author[2]{Yong-Kun Zhang}{}
\author[2,35]{Lei Zhang}{}
\author[2]{Chunfeng Zhang}{}
\author[19]{\\Rushuang Zhao}{}
\author[2]{Dejiang Zhou}{}

\address[1]{Research Center for Astronomical Computing, Zhejiang Laboratory, Hangzhou {\rm 311100}, China}
\address[2]{National Astronomical Observatories, Chinese Academy of Sciences, Beijing {\rm 100101}, China}
\address[3]{Institute for Frontiers in Astronomy and Astrophysics, Beijing Normal University,  Beijing {\rm 102206}, China}
\address[4]{State Key Laboratory of Radio Astronomy and Technology, Beijing {\rm 100101}, China}
\address[5]{University of Chinese Academy of Sciences, Beijing {\rm 100049}, China}
\address[6]{Department of Physics, University of Hong Kong, Pokfulam Road, Hong Kong, China}
\address[7]{Nevada Center for Astrophysics, University of Nevada, Las Vegas, NV 89154, USA}
\address[8]{Department of Physics and Astronomy, University of Nevada Las Vegas, Las Vegas, NV 89154, USA}
\address[9]{New Cornerstone Science Laboratory, Department of Astronomy, Tsinghua University, Beijing {\rm 100084}, China}
\address[10]{Institute for Astronomy, School of Physics, Zhejiang University, Hangzhou {\rm 310027}, China}
\address[11]{School of Astronomy and Space Science, Nanjing University, Nanjing {\rm 210023}, China}
\address[12]{Key Laboratory of Modern Astronomy and Astrophysics (Nanjing University), Ministry of Education, Nanjing {\rm 210093}, China}
\address[13]{Department of Astronomy, School of physics, Peking University, Beijing, {\rm 100871}, China}
\address[14]{Yunnan Astronomical Observatories, Chinese Academy of Sciences, Kunming {\rm 650216}, Yunnan, China}
\address[15]{Beijing Laser Acceleration Innovation Center, Huairou, Beijing, {\rm 101400}, China}
\address[16]{School of Physics, Huazhong University of Science and Technology, Wuhan, {\rm 430074}. China}
\address[17]{Purple Mountain Observatory, Chinese Academy of Sciences, Nanjing {\rm 210023}, China}
\address[18]{School of Physics and Electronic Science, Guizhou Normal University, Guiyang {\rm 550001}, China}
\address[19]{Guizhou Normal University, Guizhou Provincial Key Laboratory of Radio Astronomy and Data Processing, Guiyang {\rm 550001}, China}
\address[20]{Department of Astronomy, School of Physics and Materials Science, Guangzhou University, Guangzhou {\rm 510006}, China}
\address[21]{School of Physics and Astronomy, Beijing Normal University, Beijing {\rm 100875}, China}
\address[22]{Shanghai Astronomical Observatory, Chinese Academy of Sciences, Shanghai {\rm 200030}, China}
\address[23]{Key Laboratory of Particle Astrophysics, Institute of High Energy Physics, Chinese Academy of Sciences, Beijing {\rm 100049}, China}
\address[24]{College of Physics and Electronic Engineering, Qilu Normal University, Jinan {\rm 250200}, China}
\address[25]{School of Computer Science and Engineering, Sichuan University of Science and Engineering, Yibin {\rm 644000}, China}
\address[26]{Department of Astronomy, School of Physical Sciences, University of Science and Technology of China, Hefei {\rm 230026}, China}
\address[27]{School of Astronomy and Space Science, University of Science and Technology of China, Hefei {\rm 230026}, China}
\address[28]{Department of Astronomy, Tsinghua University, Beijing {\rm 100084}, China}
\address[29]{College of Physics and Hebei Key Laboratory of Photophysics Research and Application, Hebei Normal University, Shijiazhuang, Hebei {\rm 050024}, China}
\address[30]{Shijiazhuang Key Laboratory of Astronomy and Space Science, Hebei Normal University, Shijiazhuang, Hebei {\rm 050024}, China}
\address[31]{Institute of Astrophysics, Central China Normal University, Wuhan {\rm 430079}, China}
\address[32]{National Key Laboratory of Dark Matter Physics, School of Physics and Astronomy, Shanghai Jiao Tong University, Shanghai {\rm 200240}, China}
\address[33]{Laboratory for Laser Plasmas and Collaborative Innovation Centre of IFSA, Shanghai Jiao Tong University, Shanghai {\rm 200240}, China}
\address[34]{South-Western Institute for Astronomy Research, Key Laboratory of Survey Science of Yunnan Province, Yunnan University, Kunming, Yunnan {\rm 650500}, China}
\address[35]{Centre for Astrophysics and Supercomputing, Swinburne University of Technology, Hawthorn 3122, Australia}

\abstract{Possible periodic features in fast radio bursts (FRBs) may provide insights into their astrophysical origins. Using extensive observations from the Five-hundred-meter Aperture Spherical radio Telescope (FAST), we conduct a multi-timescale periodicity search for the exceptionally active repeater FRB~20240114A. Our analysis is based on different datasets for different timescales: for short-timescale periodicity in Time of Arrivals (TOAs), we use 57 observations from January to August 2024; for long-timescale periodicity, we employ an extended TOA dataset comprising 111 observations spanning from January 2024 to October 2025; and for burst time series analysis, we utilize individual burst data from the 57 FAST observations. We identify three candidate short-timescale periodic signals (0.673~s, 0.635~s, and 0.536~s) with significances of $3.2\sigma$--$6\sigma$, each detected in two independent observations. On longer timescales, we detect a significant $143.40\pm7.19$-day periodicity with $5.2\sigma$ significance, establishing FRB~20240114A as a periodic repeater. In burst time series, we find quasi-periodic oscillations in the few hundred Hz range ($3.4\sigma$ and $3.7\sigma$) and periodic burst trains with periods of several to tens of milliseconds ($3\sigma$--$3.9\sigma$), though these periodic features appear transient and short-lived. The detection of periodic signals at these different time scales indicates that FRB 20240114A exhibits intriguing periodic self-similar characteristics. Despite the comprehensive dataset, no definitive periodicity linked to the source's rotation is confirmed, placing stringent constraints on the intrinsic source properties and the modulation mechanisms. All data are available via the Science Data Bank.}
    
\keywords{Fast radio burst, Time series analysis, Magnetars}
\pagestyle{empty}
\maketitle

\begin{multicols}{2}
\section{Introduction} \label{sec:intro}
Fast radio bursts (FRBs), first discovered in 2007 \cite{lorimer2007bright}, are millisecond-duration, highly energetic radio transients of extragalactic origin. They have been detected at cosmological distances and are now recognized as one of the most intriguing phenomena in modern astrophysics \cite{lorimer2007bright,2013Sci...341...53T,chime/frb2021First,2022Natur.606..873N}. Since their initial discovery, the field of FRB research has grown rapidly, with extensive observational campaigns leading to the identification of hundreds of these mysterious bursts. These detections have revealed a diverse population, with some FRBs appearing to repeat sporadically while others have only been observed as singular, non-repeating events \cite{2019A&ARv..27....4P,chime/frb2021First,xu2023Blinkverse}. The identification of repeating FRBs, most notably FRB 20121102A \cite{spitler2016repeating}, has ruled out the early speculations that all FRBs originate from cataclysmic one-time events. Instead, the existence of repeaters indicates that at least some FRBs are powered by non-catastrophic engines. 

A breakthrough in the field of FRB research was marked by the detection of FRB~20200428, which was definitively associated with an X-ray burst from the Galactic magnetar SGR~J1935+2154 \cite{2020Natur.587...59B,mereghetti2020INTEGRAL,2021NatAs...5..378L}. Another targeted multi-wavelength campaign did not detect further coincident radio bursts from this source, suggesting that such FRB-SGR associations are rare and might require extreme physical conditions to produce coherent radio emission \cite{2020Natur.587...63L}. Extensive observations of repeating FRBs have revealed non-Poissonian burst clustering, long-term variations in DM and rotation measure (RM), as well as complex waiting time statistics, highlighting their highly stochastic nature and posing challenges to single rotating compact object models \cite{2021Natur.598..267L,2023ApJS..269...17H,2024SciBu..69.1020Z,2025arXiv250400391L}. Recent polarization and magnetospheric studies suggest that repeating FRBs reside in complex, evolving environments consistent with young magnetars, as indicated by rapid RM variations and strong circular polarization, although observations of FRB 20220912A, which shows high circular polarization but nearly constant RM, suggest that at least some repeaters may not be embedded in strongly magneto-ionic environments \cite{2022Natur.609..685X,2022Sci...375.1266F,2023SciA....9F6198Z,2024ApJ...974..296F,2025SCPMA..6889511F}. A successful search for pulsar-like, spin-modulated burst-to-burst temporal periodicities in FRB sources would provide a convincing observational evidence that at least some FRBs originate from compact stars. 

Some FRB sources have demonstrated distinct long-timescale periodic activity. Notable examples include FRB 20180916B, which follows a 16-days cycle of activity \cite{2020Natur.582..351C,2021Natur.596..505P,2023ApJ...956...23S}, and FRB 20121102A, which has shown signs of a much longer periodicity of approximately 160 days, alongside a shorter periodicity of around 4.6 days \cite{2020MNRAS.495.3551R,2021MNRAS.500..448C,2024ApJ...969...23L,2025A&A...693A..40B}. The recently discovered FRB 20240209A, located at the outskirts of a quiescent elliptical galaxy (z = 0.1384), exhibits a periodic activity cycle of about 126 days, further adding to the rare class of periodically repeating FRBs \cite{2025ApJ...983L..15P}. Furthermore, FRB 20220529 shows a potential periodicity of approximately 200 days in its RM, which is consistent with a binary origin, as suggested by the significant RM variation and its rapid recovery \cite{2025arXiv250510463L}. Similarly, FRB 20201124A exhibits a periodicity of 26 days in its RM, further supporting the idea that periodic RM variations may be a common feature of actively repeating FRBs in binary systems \cite{2025arXiv250506006X}. The origin of periodic activity in repeating FRBs remains debated, with proposed explanations falling into several categories. Rotation-powered models suggest the periodicity reflects the spin of isolated magnetars \cite{2021ApJ...917....2X,2024ApJ...967L..44L}. Binary models attribute the modulation to orbital dynamics in neutron star systems, including Be/X-ray binaries \cite{2020ApJ...893L..26I,2020ApJ...893L..39L,2020MNRAS.498L...1Z,2021ApJ...918L...5L,2022MNRAS.515.4217B}. Precession effects have also been proposed as a possible cause of the observed periodicity \cite{2020ApJ...893L..31Y,2020RAA....20..142T,2020ApJ...895L..30L,2020ApJ...892L..15Z,2020MNRAS.494L..64K,2020PASJ...72L...8C,2021ApJ...921..147C,2021ApJ...917...13S,2022A&A...658A.163W}. Alternative explanations involve asteroid belt interactions \cite{2020ApJ...895L...1D,2021MNRAS.508.2079V,2024ApJ...974..215D} or the repetitive collapse of the crust of an accreting strange quark star \cite{2021Innov...200152G}. The diversity of models highlights the need for further observational constraints, as many fundamental questions about the physical origins and emission mechanisms of FRBs remain open \cite{2023RvMP...95c5005Z}.

Short-timescale periodicity in FRBs provides critical constraints on their emission mechanisms and progenitors. Some bursts have shown millisecond-scale quasi-periodic structures, e.g. the 0.411 ms spacing in FRB 20201020A \cite{2023A&A...678A.149P}, although its significance remains marginal. These sub-second periodicities challenge the spin explanations due to extreme rotational requirements, favoring magnetospheric processes instead. Additionally, the detection of ultra-narrow spike components in FRB 20230607A, down to 0.3 ms, further supports a magnetospheric origin for at least some repeating FRBs \cite{2025arXiv250411173Z}. On the other hand, other studies have found no significant periodicity in repeating FRBs like FRB 20201124A, ruling out isolated magnetars or binaries with certain parameters \cite{2022RAA....22l4004N}. Recent claims of a $\sim$1.7 s period in FRB 20201124A \cite{2025arXiv250312013D}, if confirmed, could indicate a pulsar origin, while low-significance periodicity candidates such as the 7.267 ms signal in FRB 20230708A \cite{2025MNRAS.536.3220D} highlight the challenges in distinguishing true periodicity from burst microstructure. These controversial results highlight the need for higher-cadence and higher time resolution observations to clarify the nature of short-timescale FRB variability.

\begin{figure*}
    \centering
    \includegraphics[scale=0.373]{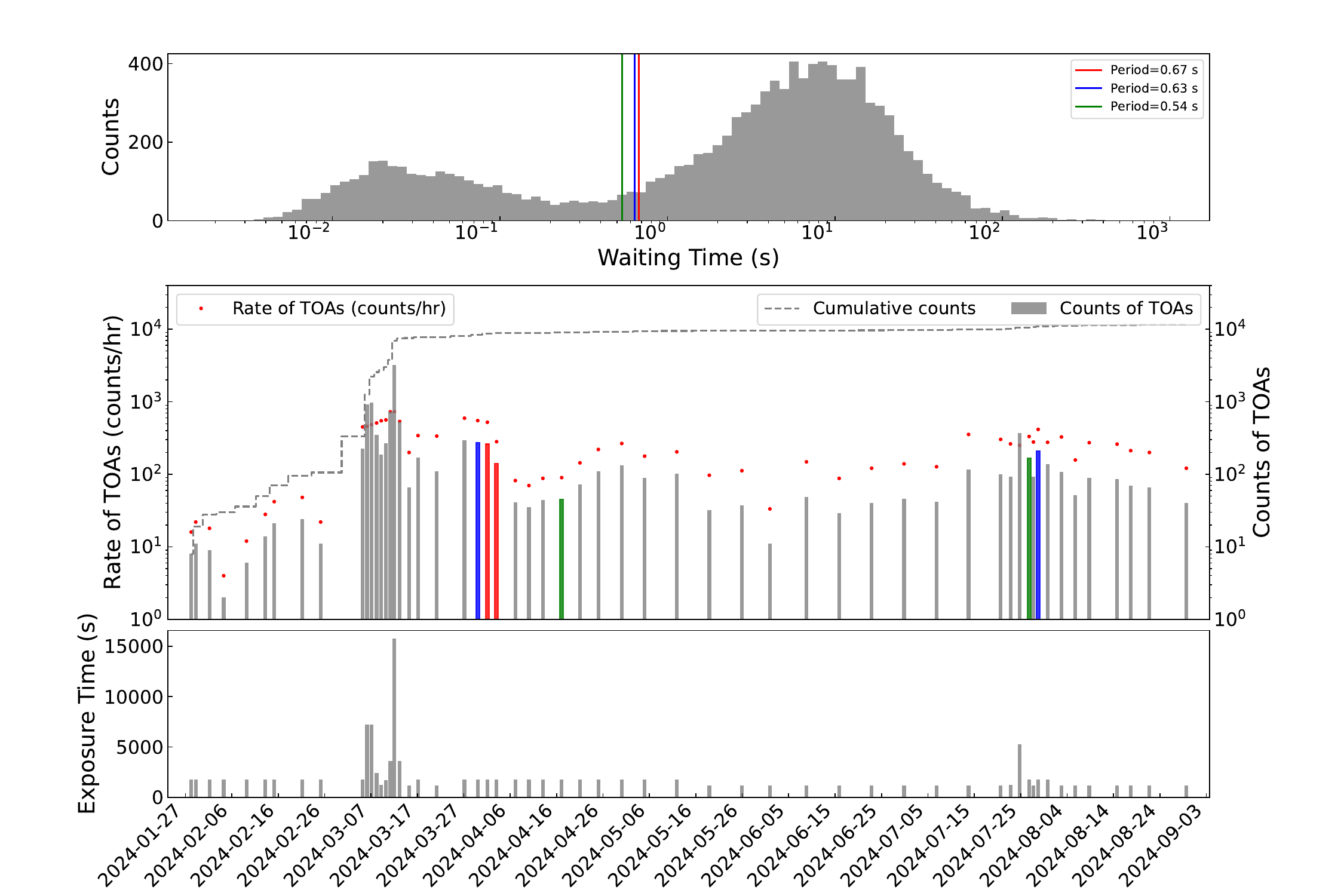}
    \caption{The 57 FAST observations of FRB 20240114A from January 28 to August 29, 2024 (UT), show the distribution of waiting times between bursts (top panel), burst counts (bars in the middle panel), burst rates (red dots in the middle panel), cumulative burst counts (dashed line in the middle panel), and exposure times (bars in the bottom panel). In the top panel, vertical lines of different colors represent the period values of three candidates identified in the TDA short-timescale search. The bar charts with the same colors in the middle panel indicate that there are candidate periodic signals with similar parameters in the corresponding observations (see Section \ref{sec:period_search}).}
    \label{fig:toas}
\end{figure*}

This study systematically analyzes the periodicity of FRB 20240114A, an extremely active FRB. The source was discovered by the Canadian Hydrogen Intensity Mapping Experiment (CHIME) in January 2024, and since then, it has been observed by multiple telescopes \cite{2024ATel16420....1S, 2024MNRAS.533.3174T, 2024arXiv241010172X, 2025arXiv250403569H}. The source has been continuously monitored by the FRB key project of the Five-hundred-meter Aperture Spherical radio Telescope (FAST) \cite{2011IJMPD..20..989N}. For the analysis presented in this work, we utilize two complementary datasets: (1) a core sample spanning from January 28 to August 29, 2024 (UT), comprising 11,553 burst events collected over 57 observation sessions (see Figure \ref{fig:toas}), which is used for short-timescale periodicity searches and most other analyses; and (2) an extended sample incorporating additional observations up to October 1, 2025 (UT), employed exclusively for long-timescale periodicity searches to enhance sensitivity to potential multi-month activity cycles (see Figure \ref{fig:long_results}A). All bursts were recorded using the 19-beam L-band receiver of FAST \cite{2018IMMag..19..112L}, covering a frequency range of 1.0--1.5~GHz with a time resolution of 49.152~$\mu$s. The processed data corresponding to the core sample are publicly accessible via the Science Data Bank\footnote{\url{https://doi.org/10.57760/sciencedb.Fastro.00030}}. Other studies based on the core dataset include \cite{zhang2025a, zhang2025b}. By combining both the core and extended datasets, this study presents the largest-sample investigation to date into multiple-timescale periodicities and quasi-periodic oscillations (QPOs) from a single FRB source.

This paper is organized as follows. In Section \ref{sec:period_search}, we search for periodic signals in the time-of-arrival (TOA) data of FRB 20240114A to explore possible links to magnetar rotation, binary orbits, and precession scenarios. In Section \ref{sec:ts_search}, we mainly investigate QPOs in the burst time series, which provide crucial insights into the underlying radiation mechanisms. Section \ref{sec:discussion} discusses the implications of our findings, and Section \ref{sec:summary} summarizes the main conclusions of this study.

\section{Searching for Periodic Signals in TOA data}\label{sec:period_search}
The primary TOAs analyzed in this paper, which are used for most analyses including the short-timescale periodicity search, are derived from 57 FAST observations of FRB 20240114A between January 28, 2024, and August 29, 2024 (UT). Figure \ref{fig:toas} shows the time distribution of bursts across these 57 observations, as well as the distribution of waiting times between consecutive TOAs within this dataset. 

To enhance the sensitivity of the long-timescale periodicity search to multi-month activity cycles, we specially extend the dataset to include TOAs from observations up to October 1, 2025 (UT). While the short-timescale analyses rely on the original dataset for high-precision single-pulse measurements suitable for detecting minute-to-hour periodicities, the extended dataset provides reliable TOAs and burst rates appropriate for detecting long-term activity cycles. For more details on the data processing of this source, including burst searching, dispersion correction, and TOA definition, please refer to \cite{zhang2025a}. The TOAs were converted to the solar system barycenter using \texttt{PINT} \cite{2021ApJ...911...45L, 2024ApJ...971..150S}.

\subsection{Method}
\subsubsection{Short-Timescale Periodicity Search}
Short-timescale searches refer to conducting periodic searches on the obtained TOA data, covering timescales ranging from several tens of milliseconds to tens of minutes. A dual analysis approach was employed for each continuous observation dataset. This involved first performing a wide-range scan of frequency $f$ and its first derivative $\dot{f}$ using the Time-Differencing Algorithm (TDA), followed by precise period determination through the $Z^2$ statistic.

\begin{table*}
\centering
\scriptsize
\caption{Candidate periodic signals detected by the TDA method and refined using the $Z_n^2$ statistic. The table lists the candidate number, observation date (MJD), parameters from the TDA detection (frequency $f$, derivative $\dot{f}$, period $P$, derivative $\dot{P}$, significance $\sigma$), and the refined parameters from the $Z_n^2$ statistic.}\label{tab:short_TDA}
\begin{tabular}{ccccccccccc}
\toprule
\multirow{2}{*}{No.} & \multirow{2}{*}{MJD} & \multicolumn{5}{c}{TDA Detection} & \multicolumn{4}{c}{$Z_n^2$ Refinement} \\
\cmidrule(lr){3-7} \cmidrule(lr){8-11}
 & & $f$ & $\dot{f}$ & $P$ & $\dot{P}$ & $\sigma$ & $f$ & $\dot{f}$ & $P$ & $\dot{P}$ \\
  & & Hz & $\times 10^{-7}$ Hz/s & s & $\times 10^{-7}$ s/s &  & Hz & $\times 10^{-7}$ Hz/s & s & $\times 10^{-7}$ s/s \\
\midrule
\multirow{2}{*}{1} & 60401 & 1.48677(58) & -14.322(78) & 0.67260(26) & 6.479(36) & 4.9 & 1.4867871(30) & -14.319070(29) & 0.6725912(13) & 6.477646(29) \\
& 60403 & 1.48482(58) & -1.398(78) & 0.67348(26) & 0.634(35) & 4.2 & 1.4849567(30) & -1.3992047(28) & 0.6734203(13) & 0.6345322(28) \\
\hline
\multirow{2}{*}{2} & 60398 & 1.57588(59) & 5.138(85) & 0.63457(24) & -2.069(34) & 3.2 & 1.5761418(32) & 5.133988(10) & 0.6344607(13) & -2.0666374(93) \\
& 60519 & 1.57392(59) & 7.686(85) & 0.63536(24) & -3.103(34) & 6.0 & 1.5737813(32) & 7.692812(15) & 0.6354123(13) & -3.105964(14) \\
\hline
\multirow{2}{*}{3} & 60417 & 1.86477(58) & -2.729(98) & 0.53626(17) & 0.785(28) & 3.6 & 1.8648264(37) & -2.7264407(55) & 0.5362429(11) & 0.7840057(35) \\
& 60517 & 1.86652(58) & 8.549(97) & 0.53576(17) & -2.454(28) & 3.5 & 1.8665689(37) & 8.557654(17) & 0.5357423(11) & -2.456216(11) \\
\bottomrule
\end{tabular}
\end{table*}

The TDA has been widely applied in periodicity detection for high-energy gamma-ray pulsars \cite{2006ApJ...652L..49A,2008ApJ...680..620Z,2008Sci...322.1218A,2010ApJS..187..460A,2023ApJ...958..191S}. In particular, Atwood et al. (\cite{2006ApJ...652L..49A}) demonstrated that TDA can achieve comparable sensitivity to FFT-based searches while being over an order of magnitude faster for sparse TOA data, making it especially suitable for FRBs, whose bursts are typically sparsely and irregularly distributed over the observation period. The TDA evaluates periodicity by analyzing time differences between events across a sliding time window. The statistical quantity searched by this method is \cite{2006ApJ...652L..49A}
\begin{equation} 
D_l^{N_w} = \sum_{m=k+1}^{M}\sum_{k=0}^{N-1}a_k a_m e^{-i2\pi l(m-k)/N}, 
\end{equation}
where $a_k$ represents the number of TOAs in the kth time sample and $N$ is the number of time samples in the overall viewing period $T_v$. The upper limit for the outer summation is given by $M = {\rm min}(N-1, k+N/N_{w})$, where $N/N_{w}=NT_{\rm win}/T_v$ represents the number of time samples within the window defined by $T_{\rm win}$. When $|{\rm Re}(D_l^{N_w})|$ exceeds a predetermined threshold, the corresponding frequency is identified as a candidate.

Our search simultaneously covers both frequency $f$ and its derivative $\dot{f}$, accounting not only for possible spin-down effects of the FRB source but also for potential binary system scenarios, provided that $T_{v} \leq 10P_{\rm b}$, where $P_{\rm b}$ is the orbital period of the FRB source \cite{2003ApJ...589..911R}. The correction for $\dot{f}$ can be determined through
\begin{equation}
t_i = \tilde{t_i} + \frac{1}{2} \frac{\dot{f}}{f} \tilde{t_i}^2,
\end{equation}
where $\tilde{t_i}$ and $t_i$ represent the original and corrected TOAs respectively. In our search, the frequency range extends from the lower bound determined by the observation duration up to 32 Hz, with two harmonics combined to enhance the signal-to-noise ratio. The step size for searching $\frac{\dot{f}}{f}$ is given by
\begin{equation}
    d\left(\frac{\dot{f}}{f}\right) = \frac{1}{T_{\rm win}T_{v}f_{\rm max}},
\end{equation}
where $f_{\rm max}$ represents the maximum searched frequency. This parameter space covers plausible physical scenarios for FRB sources. $T_{\rm win}$ balances computational efficiency and frequency resolution: for observations $<$ 1800 s, $T_{\rm win}=T_{v}$; for observations $\geq$ 1800 s, $T_{\rm win}$ is fixed at 1800 s. Previous studies by the reference \cite{2022NatCo..13.4382W} suggest that repeating FRBs might originate from magnetars in binary systems with Be stars possessing disk structures. The range of $\dot{f}/f$ explored spans physically plausible regimes for both spin-down and binary motion. In a circular binary orbit, this quantity relates to the orbital period $P_{\rm b}$ and companion mass $M_{\rm c}$ via 
\begin{equation}
\left|\frac{\dot{f}}{f}\right|_{\rm max} = \frac{a_{\rm max}}{c}=\frac{(2\pi)^{\frac{4}{3}}G^{\frac{1}{3}}M_c}{(M_p+M_c)^{\frac{2}{3}}}P_b^{-\frac{4}{3}}c^{-1}.\label{eq:f1_f0_Cir}
\end{equation}
Following the reference \cite{2022RAA....22l4004N}, we adopt an upper limit of $M_c = 100 M_\odot$ to represent an extreme massive companion. Under this assumption, $\left(\dot{f}/f\right)_{\max} = 10^{-6}\ \mathrm{s^{-1}}$, which corresponds to an orbital period of $P_b \sim 8.2$ hr. We also consider potential spin-down effects of the FRB source. For magnetars, the magnetic field strength of the surface $B_s$ is related to the spin period $P$ and its derivative $\dot{P}$ through $B_s = 3.2 \times 10^{19}\,\mathrm{G} \sqrt{P \dot{P}}$. Assuming a magnetar surface field of $10^{15}\,\mathrm{G}$, the intrinsic frequency evolution yields $\frac{\dot{f}}{f} \sim -9.77 \times 10^{-10}f^2\ \mathrm{s^{-1}}$. Considering the known spin frequency of magnetars\footnote{https://www.physics.mcgill.ca/~pulsar/magnetar/main}, the value of $\frac{\dot{f}}{f}$ ranges from $-9.5 \times 10^{-9}\ \mathrm{s^{-1}}$ to $-7.13 \times 10^{-12}\ \mathrm{s^{-1}}$.

Our search strategy includes three complementary approaches:
\begin{enumerate}
    \item Dedicated search: The TOAs from each single continuous observation are analyzed separately,  covering frequency $f$ and its derivative $\dot{f}$ with $\frac{\dot{f}}{f}$ ranging from $-10^{-6}$ to $10^{-6}\,\mathrm{s}^{-1}$.
    \item Joint search across days: All TOAs data from multiple observations are searched collectively, with a refined $\frac{\dot{f}}{f}$ range of $-10^{-8}$ to $0\,\mathrm{s}^{-1}$ to focus on long-term spin-down trends.
    \item Clustered search: Since a single \( \dot{f} \) may not fully capture the frequency evolution over long timescales, the TOAs are divided into segments (each $\leq$ 3 days) and analyzed independently. For each segment, \( \frac{\dot{f}}{f} \) is constrained to \(-10^{-8}\) to \(0 \, \mathrm{s}^{-1}\), allowing for a more precise characterization of the spin-down behavior within shorter intervals.  
\end{enumerate}
The search range is designed to account for both long-term spin-down evolution and potential binary orbital dynamics, thereby ensuring sensitivity to strongly magnetized magnetars in either scenario.

Candidate signals are identified by filtering those exceeding a threshold, based on fitting the complementary cumulative distribution function (CCDF) of the TDA spectrum to an exponential function $\mathrm{CCDF}(x) = A \cdot e^{-Bx}$, as described in the reference \cite{2008ApJ...680..620Z}. The $P$-value for each power $x$ is calculated from the CCDF to assess its significance. Given that the search over frequency and its first derivative is a classic multiple testing problem, we control the false alarm rate by multiplying the obtained $P$-values by a correction factor $C$, where $C$ equals the product of the number of searched frequencies and the number of searched first-derivative values (ranging symmetrically in both positive and negative directions around zero). The corrected $P$-value is given by $P_{\rm corrected} = \min(1, P \cdot C)$. This method imposes more stringent penalties on $P$-values corresponding to first-derivative values farther from zero, effectively suppressing false alarms while preserving potential true signals. Candidates are those with $P_{\rm corrected} \leq 2.7 \times 10^{-3}$, corresponding to a 3$\sigma$ threshold. We refine each candidate $(f, \dot{f})$ using the $Z^2_n$ statistic \cite{1983A&A...128..245B}
\begin{equation}
Z^2_n = \frac{2}{N}\sum_{k=1}^{n}\left[\left(\sum_{j=1}^{N}\cos k\phi_j\right)^2 + \left(\sum_{j=1}^{N}\sin k\phi_j\right)^2\right],
\end{equation}
where N is the number of TOAs and $\phi_j = 2\pi(f t_j + \frac{1}{2}\dot{f} t_j^2)$. We evaluate the significance of each candidate by computing $Z^2_n$ over a fine grid around the TDA-derived $(f, \dot{f})$, incorporating the first $n=4$ harmonics to balance sensitivity and computational cost. This two-stage approach, which combines a wide search using the TDA with subsequent $Z^2_n$ refinement, ensures efficient and reliable detection of weak periodic signals in sparse TOAs. A demonstration of this TDA and $Z_n^2$ pipeline on simulated TOA data is provided in \ref{appendix:TDA}.
\begin{figure*}
    \centering
    \includegraphics[scale=0.3]{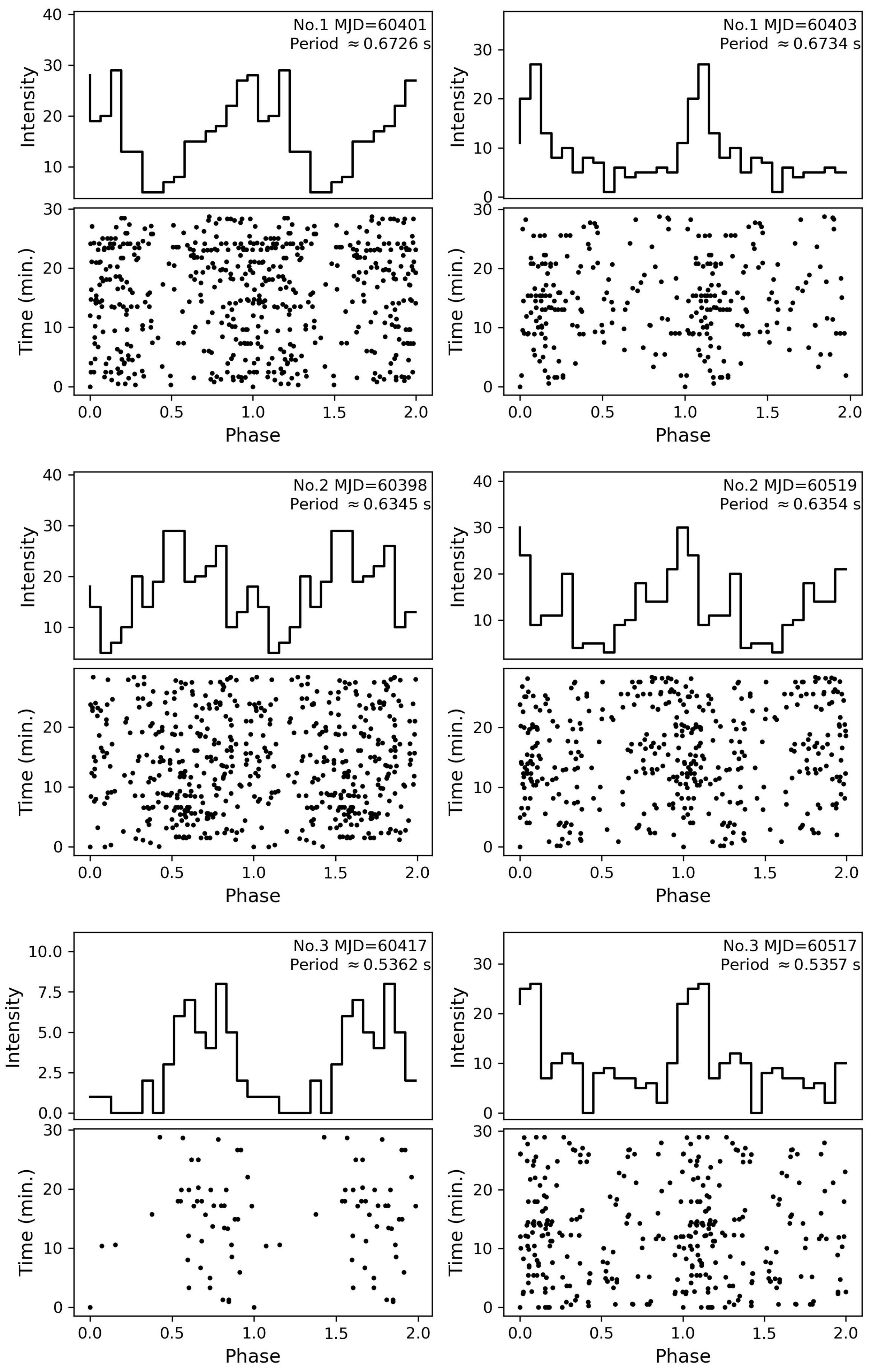}
    \caption{Folding results of the short-timescale periodicity candidates. Each row corresponds to one candidate group, with the left and right panels showing similar periods observed on different observation dates (see Table~\ref{tab:short_TDA}). Top of each panel indicates the candidate period \(P\) in seconds. For each candidate, the top panel displays the pulse profile folded with the corresponding period after correcting for the period derivative, and the bottom panel shows the distribution of individual TOAs in phase.}
    \label{fig:shortf}
\end{figure*}

\subsubsection{Long-Timescale Periodicity Search}
The long-timescale search aims to detect periodicity on timescales of the order of days, corresponding to the activity cycle of the FRB. The primary analysis employs Lomb-Scargle periodogram (LSP) analysis, with epoch folding used for verification of candidate signals.

For the long-timescale search, we include all TOAs from observations between 2024-01-28 and 2025-10-01 (UT) to maximize the time span and improve sensitivity to multi-month periods. The extended dataset provides reliable TOAs and burst rates suitable for long-timescale analyses. This approach ensures both high data quality and optimal sensitivity for detecting multi-month periodicity. The complete FAST observational log used for this analysis is provided in \ref{app:observing_log}.

For the LSP analysis, the weighted mean TOAs from each observation are used as the time series, with weights proportional to the daily burst rate.
The search is conducted over frequencies corresponding to periods from $P_{\rm min}=1.5$~days to $P_{\rm max}=T_{\rm span}/2 \approx 306.16$~days, where $T_{\rm span} = 612.33$~days is the total time span of the observations.
The frequency grid is sampled with 100 points per independent peak, resulting in a frequency step of
\begin{equation}
df = \frac{f_{\rm max}-f_{\rm min}}{N_{\rm samp} \cdot N_{\rm peak}},
\end{equation}
where $f_{\rm min}=1/P_{\rm max}$, $f_{\rm max}=1/P_{\rm min}$, $N_{\rm samp}=100$ is the number of samples per peak, and $N_{\rm peak}$ is the estimated number of independent peaks within the frequency range.
This ensures that narrow peaks can be resolved while covering the full period range of interest.
The LSP power is computed using the standard Lomb-Scargle formulation.
The significance of detected peaks is quantified using the false alarm probability (FAP) (\cite{2018ApJS..236...16V}):
\begin{equation}
{\rm FAP}(z) \approx 1- \left[P_{\rm single}(z)\right]^{N_{\rm eff}},
\end{equation}
where
\begin{equation}
P_{\rm single}(z) = (1-z)^{\frac{N-1}{2}},
\end{equation}
$N$ is the number of data points, and $N_{\rm eff}=f_{\rm max} T_{\rm span}$ is the effective number of independent frequencies.

Candidate periods are selected based on the FAP significance level, with a threshold corresponding to $>3\sigma$ confidence (FAP $< 0.0027$). For verification of these LSP-detected candidates, we employ epoch folding to examine the phase profile. 

Folded profiles can be modeled with one or more Gaussian functions plus a constant baseline:
\begin{equation}
f(x) = \sum_{k=1}^{N} A_k \exp\left[-\frac{(x-\mu_k)^2}{2\sigma_k^2}\right] + C,
\end{equation}
where $N$ is the number of Gaussian components, $A_k$ the amplitude of the $k$-th component, $\mu_k$ its phase center, $\sigma_k$ its width, and $C$ the constant baseline. Exposure correction is applied when necessary, and the profile may be repeated or smoothed to facilitate visualization and comparison.
\begin{figure*}
    \centering
    \includegraphics[scale=0.18]{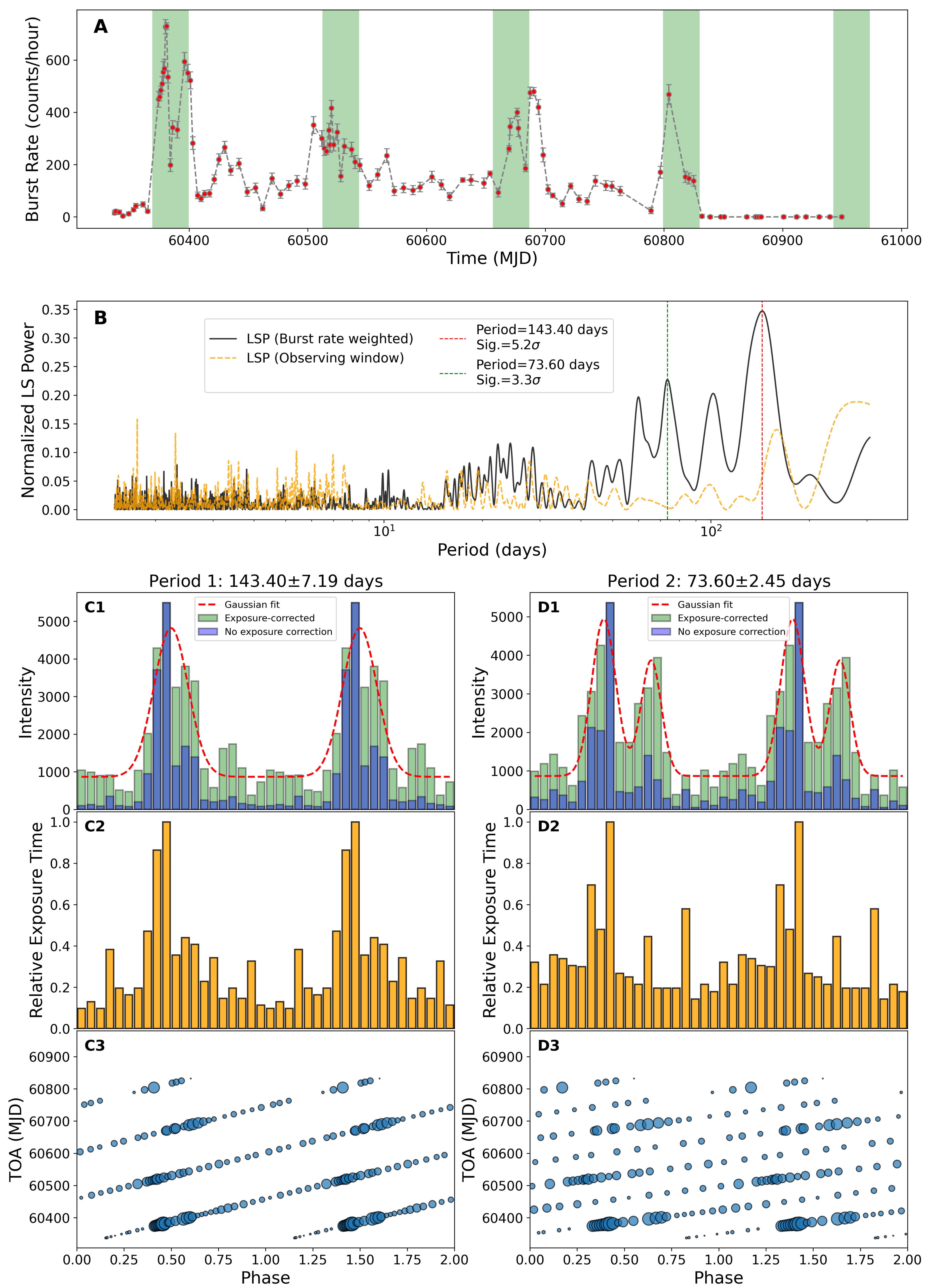}
    \caption{Long-timescale periodicity search results for FRB 20240114A. The analysis uses all TOAs from observations between 2024-01-28 and 2025-10-01 ($T_{\rm span} = 612.33$ days), searching periods from 1.5 days to $T_{\rm span}/2 \approx 306.16$ days. Candidate periods were selected at $>3\sigma$ confidence (FAP $< 0.0027$) and verified through epoch folding with Gaussian profile modeling. \textbf{A}: Burst rate evolution with time, showing the count rate (counts/hour) versus MJD. The green-shaded regions mark the predicted enhanced active phases. These are derived from a phase-folding analysis using the 143.40-day period, with a Gaussian fit (panel C1) applied to identify the timing and duration of the expected enhanced activity windows. \textbf{B}: Lomb-Scargle periodogram showing normalized power versus period. The black solid line represents the burst rate-weighted analysis, while the orange dashed line shows the observing window function. Two significant periods are identified: Period 1 = $143.40 \pm 7.19$ days (red dashed line, $5.2\sigma$ significance) and Period 2 = $73.60 \pm 2.45$ days (green dashed line, $3.3\sigma$ significance). \textbf{C1-C3}: Phase-folded analysis for Period 1. \textbf{C1}: Intensity profile showing exposure-corrected (green) and uncorrected (blue) rates with Gaussian fit (red dashed), from which the enhanced active window timing and duration are determined. \textbf{C2}: Relative exposure time distribution across phase bins. \textbf{C3}: TOA versus phase, with point sizes scaled by burst rate weights. \textbf{D1-D3}: Similar to C1-C3 but for Period 2. Based on the analysis results, Period 2 is identified as a harmonic of Period 1.}
    \label{fig:long_results}
\end{figure*}

\subsection{Results}
\subsubsection{Short-Timescale Periodicity}
We first performed a joint search of all observational data, which did not yield any candidates meeting our selection criteria. Subsequently, we conducted an analysis of individual observations by grouping those taken within 2–3-day intervals. This clustering approach resulted in 14 distinct groups of TOA data being analyzed, none of which produced candidates that satisfied our criteria. The individual observation analysis revealed several candidates that met our selection criteria. We require each observation to include at least three TOAs to be included in the search, which resulted in the exclusion of one observation from MJD 60344 out of the total 57 observational datasets. We detected 2765 candidates exceeding the $3\sigma$ significance level after multiple testing correction. Given the wide search parameter space and sparse TOA distribution, such a number of apparent candidates is statistically expected. Therefore, we imposed stringent criteria to isolate the most reliable signals: cross-epoch recurrence and minimum $P$-value within frequency clusters. First, for candidates with similar $f$ (where similarity is defined as $\Delta f / f < 10^{-3}$) but different $\dot{f}$, only the candidate with the smallest $P$-value was retained. Subsequently, we retained candidates that reappeared in at least two separate observation epochs. This rigorous process ultimately identified three candidates exhibiting cross-epoch recurrence. The dates on which these signals appeared are marked in the same color in Figure \ref{fig:toas}. 

For precise parameter determination, we performed a refined search in their $f$-$\dot{f}$ parameter space using the $Z_n^2$ statistic. Specifically, for each candidate signal, we searched for 16,000 combinations of $f$ and $\dot{f}$ centered on the values determined by the TDA, with step sizes chosen such that the maximum phase drift over the entire continuous observation period would be less than 0.01. The $Z_n^2$ analysis led to the identification of three distinct candidates: one with a period of $\sim$0.673 s (detected in observations 60401 and 60403), another with a period of $\sim$0.635 s (detected in observations 60398 and 60519), and a third with a period of $\sim$0.536 s (detected in observations 60417 and 60517). Table~\ref{tab:short_TDA} presents the sifted candidates identified by the TDA through rigorous selection criteria, alongside their refined parameters obtained via precision analysis with the $Z_n^2$ statistic. Figure \ref{fig:shortf} displays the profile and the corresponding time-phase diagram using these optimized parameters. 

\subsubsection{Long-Timescale Periodicity}
After conducting the periodicity search on the data from 2024-01-28 to 2025-10-01 (UT), we obtained two periodic candidates. Figure~\ref{fig:long_results}A shows the burst rate evolution with time, displaying the count rate (counts/hour) versus MJD. The Lomb-Scargle periodogram (Figure~\ref{fig:long_results}B) identified two prominent peaks exceeding the \( 3\sigma \) confidence threshold (FAP $<$ 0.0027). The burst rate-weighted periodogram (black solid line) shows clear detections at both periods, while the window function (orange dashed line) confirms these are not artifacts of the observational sampling.

The primary period, designated as Period 1, was detected at \( 143.40 \pm 7.19 \) days with a significance of \( 5.2\sigma \). The secondary period, Period 2, was found at \( 73.60 \pm 2.45 \) days with \( 3.3\sigma \) significance. The period uncertainties were estimated using the relation \( \Delta P = P \cdot W / T_{\rm span} \), where \( P \) is the period in days, \( W \) is the duration of the enhanced activity window in days, and \( T_{\rm span} = 612.33 \) days is the total observational time span.

Figure~\ref{fig:long_results}C1--C3 present the phase-folded analysis for Period 1. Figure~\ref{fig:long_results}C1 shows the intensity profile with both exposure-corrected (green) and uncorrected (blue) rates, along with the Gaussian fit (red dashed). The profile is well-described by a single Gaussian component:
\begin{equation}
f_1(x) = 3951.20 \exp\left[-\frac{(x-0.499)^2}{2 \times (0.0909)^2}\right] + 867.28.
\end{equation}
From this Gaussian fit, the active window timing and duration are determined. Based on these results, the green shaded regions in Figure~\ref{fig:long_results}A mark the predicted active phases, showing the timing and duration of expected burst windows. Figure~\ref{fig:long_results}C2 displays the relative exposure time distribution across phase bins, showing relatively uniform coverage. Figure~\ref{fig:long_results}C3 presents the time of arrival (TOA) versus phase diagram, where point sizes are scaled by burst rate weights, demonstrating coherent phase folding over the entire observational baseline.

Figure~\ref{fig:long_results}D1--D3 show similar analyses for Period 2. Figure~\ref{fig:long_results}D1 reveals a double-peaked structure in the intensity profile, requiring a double-Gaussian model:
\begin{equation}
\begin{aligned}
f_2(x) = & 4072.09 \exp\left[-\frac{(x-0.391)^2}{2 \times (0.0635)^2}\right] + \\
& 3000.12 \exp\left[-\frac{(x-0.642)^2}{2 \times (0.0542)^2}\right] + 866.29.
\end{aligned}
\end{equation}
Figure~\ref{fig:long_results}D2 shows the exposure time distribution, while Figure~\ref{fig:long_results}D3 displays the TOA-phase relation with burst rate weighting.

Notably, Period 2 (\( 73.60 \pm 2.45 \) days) is consistent with being the first harmonic of Period 1 (\( 143.40 \pm 7.19 \) days), with the ratio \( P_1/P_2 = 1.95 \pm 0.12 \) compatible with a 2:1 relationship within uncertainties. The phase-folded diagrams in Figure~\ref{fig:long_results}C3 and Figure~\ref{fig:long_results}D3 support the interpretation that Period 1 (143.40 days) is the primary period, with Period 2 arising as its harmonic. Therefore, all subsequent discussion will focus solely on Period 1.

\section{searching for (quasi-)periodic signals in burst time series data}\label{sec:ts_search}
\subsection{Method}\label{sec:TS_method}
To detect periodic signals and QPOs in FRB time series, we employed two complementary analysis methods, focusing respectively on the structures within individual bursts and on the temporal patterns among burst clusters. These approaches differ primarily due to the distinct properties of the data. Burst time series typically exhibit complex non-stationary structures that resist decomposition into independent subcomponents. Consequently, frequency-domain analysis based on the Fast Fourier Transform (FFT) and Bayesian inference is required. Since the individual bursts within a burst cluster are relatively independent, measurable parameters such as subpulse widths and amplitudes can be utilized. These parameters, combined with the autocorrelation function (ACF) and phase randomization methods, facilitates the evaluation of the significance of periodic signals, thereby enabling the effective periodicity searches.

\begin{figure*}
\centering
\includegraphics[scale=0.38]{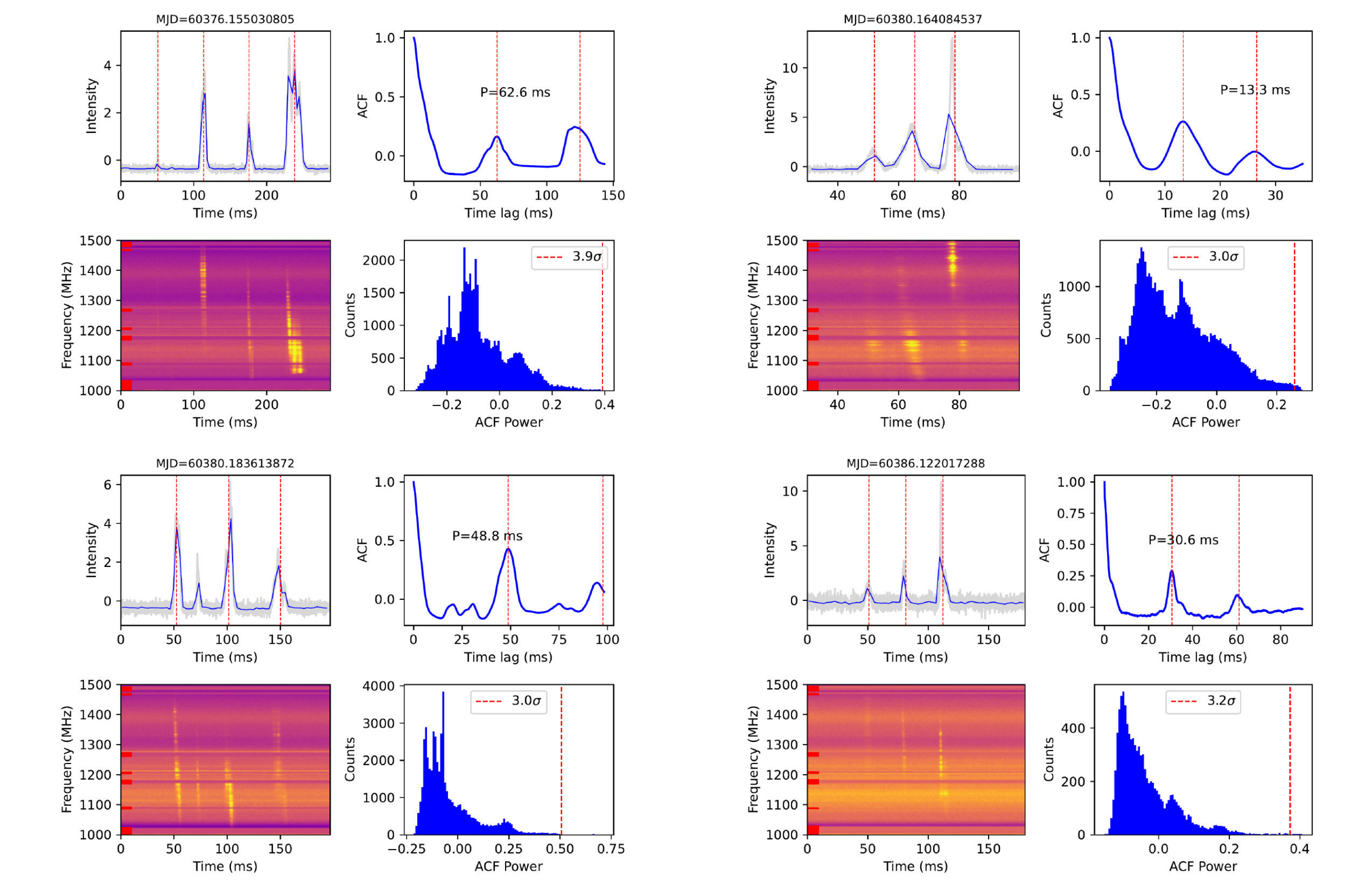}
\caption{Four quasi-periodic candidates identified from the burst time series using the ACF method, each with a significance exceeding $3\sigma$. Each candidate consists of four subplots: the top-left subplot shows the burst time series, the bottom-left subplot shows the corresponding dynamic spectrum, the top-right subplot shows the ACF of the burst time series, and the bottom-right subplot shows the simulated statistic used to calculate the probability of generating this ACF pattern. The red line in the burst time series represents the periodic intervals determined by the ACF analysis, while the red line in the simulated statistic corresponds to the observed value.}
\label{fig:ACF}
\end{figure*}

For periodicity detection of bursts, we adopted the Bayesian framework proposed by the reference \cite{2013ApJ...768...87H}. The method begins by computing the periodogram of burst time series via the FFT, then fits a null hypothesis model to obtain a maximum a posteriori (MAP) parameters. We calculate residuals $R_j = 2I_j/S_j$, where $I_j$ represents the observed periodogram and $S_j$ denotes the null model's prediction. Markov Chain Monte Carlo (MCMC) sampling generates simulated periodograms, establishing a distribution of maximum residuals against which observed residuals are compared for significance assessment. In order to detect QPOs, we further introduce a nested alternative hypothesis model, computing the likelihood ratio test (LRT) statistic $T_{\mathrm{LRT}}^{\mathrm{obs}}$ and determining its significance through simulated periodograms.

For periodicity detection of burst clusters, we employed the ACF method proposed by the reference \cite{2024NatAs...8..230K}. The method first requires determining an appropriate time window. By analyzing the distribution of burst waiting times, we found that the distribution exhibits a bimodal structure with a transition around 0.2 seconds. Based on this, bursts with intervals shorter than 0.2 seconds are clustered into the same time series. Periodic signals in ACF analysis manifest as equally spaced local maxima. We quantify the strength of periodicity using the statistic $\mathcal{P} = \sum \rho(P_\mu \cdot i)$, where $\rho$ denotes the power of the ACF, and $P_\mu \cdot i$ represents integer multiples of the period $P_\mu$. The original subpulse counts, widths, and amplitudes are preserved, while uniformly distributed perturbations are incorporated at periodic intervals. The simulated sequences then undergo the same ACF analysis to generate a distribution for comparison with the observed values, thereby obtaining the $P$-value.

The periodicity detection method for individual bursts circumvents the challenges of decomposing complex signals into subcomponents. Conversely, the method for burst clusters leverages the timing characteristics of discrete events to facilitate the assessment of periodic significance. For details of the methodology, see references \cite{2022Natur.607..256C,2024NatAs...8..230K} for the ACF-based analysis and \ref{sec:appendix} for the implementation of the FFT-based approach.

\begin{figure*}
    \centering
    \includegraphics[scale=0.7]{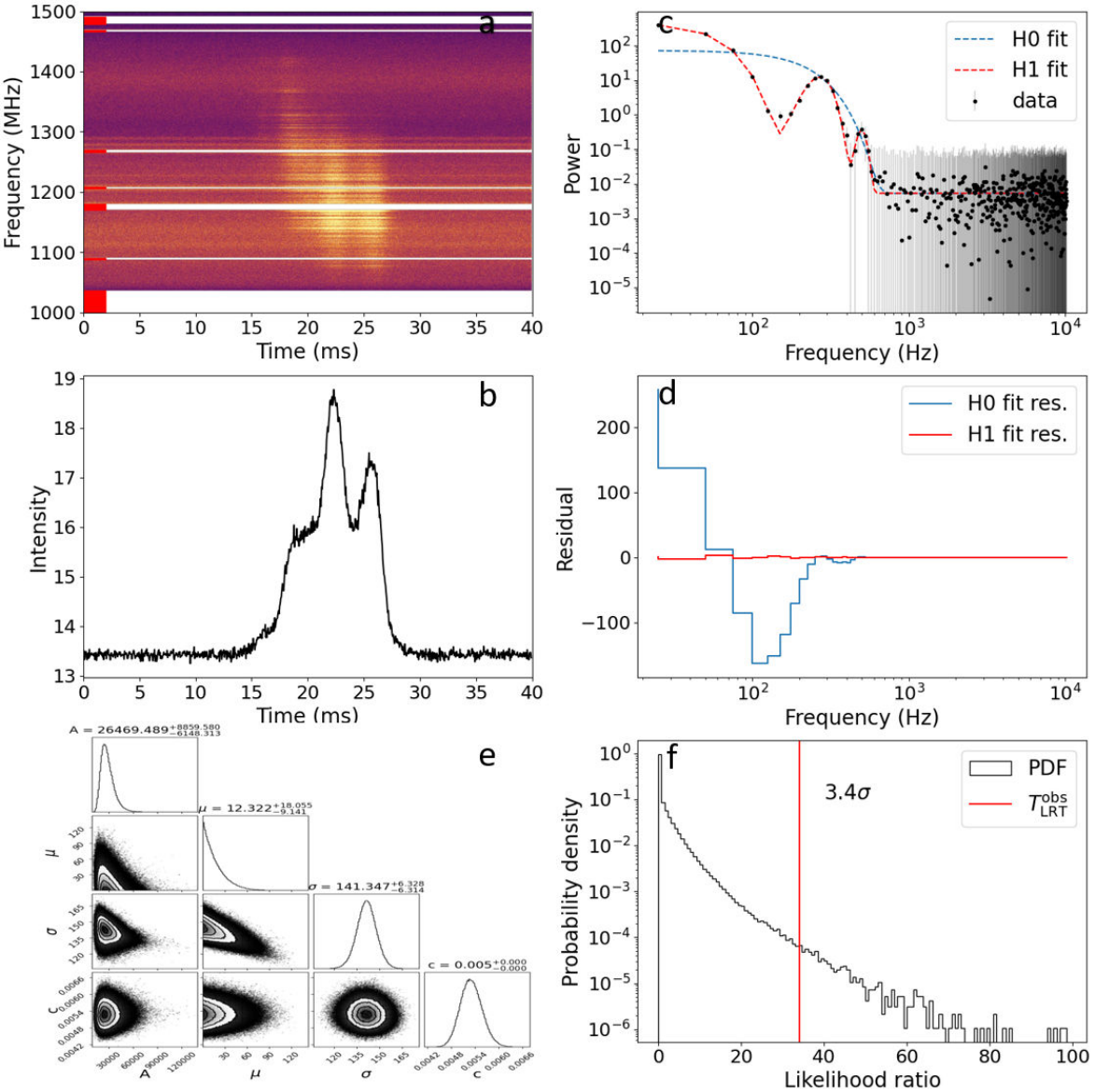}
    \caption{Analysis of an FRB burst containing a QPO component using the FFT-based Bayesian inference method. The six panels illustrate key steps in data processing and model evaluation. (a) Frequency–time dynamic spectrum of the burst, dedispersed using a dispersion measure of 528.5\,pc\,cm$^{-3}$, with RFI-contaminated channels masked; (b) Time series obtained by frequency integration of the dynamic spectrum, used as input for the periodicity analysis; (c) Periodogram of the time series. The blue dashed line represents the background power spectrum modeled under the null hypothesis $H_0$ (no QPO component), while the red dashed line corresponds to the alternative hypothesis $H_1$ (with a QPO component). A prominent excess of power is seen at a specific frequency. The best-fit parameters for the $H_1$ model are listed in Table~\ref{tab:QPO_pars}, entry No. 1. (d) Residuals of the periodogram under both $H_0$ and $H_1$, used to quantify deviations from the respective model fits; (e) Posterior distribution of model parameters obtained via MCMC sampling under $H_0$, used to generate simulated periodograms; (f) Distribution of likelihood ratio test statistics computed from the simulated periodograms. The red line marks the observed test statistic $T_\mathrm{LRT}^{\mathrm{obs}}$, which quantifies the significance of the QPO component. The observed significance of this QPO, as inferred from panel (f), is $3.4\sigma$.}
    \label{fig:QPO1}
\end{figure*}

\subsection{Results}\label{sec:QPO_results}
We applied both the ACF and the FFT methods to detect periodic signals and QPOs in the burst time series. 

For the ACF analysis, we clustered bursts using a 0.2-second time window, resulting in 516 time series. The ACF was calculated for each of these time series, and the minimum lag of the significant peaks was identified as the candidate period $P_\mu$. The significance of these periodicities was assessed using the methodology outlined in Section \ref{sec:TS_method}. We identified 4 candidates with significance greater than $3\sigma$. These burst time series, their dynamic spectra, ACFs, and distributions of simulated statistics are shown in Figure \ref{fig:ACF}. These time series exhibit relatively few peaks, with those peaks spaced at regular intervals. Most of these candidates exhibit 3 peaks, with one candidate showing 4 peaks, resulting in a high significance of up to $3.9\sigma$. In Figure~\ref{fig:ACF_2} in \ref{appendix:QPO}, we also list 7 additional candidates with significance lower than $3\sigma$ yet still showing some evidence of periodicity. Notably, some of these candidates match the period well up to 3 peaks, although the agreement breaks down after the 4th peak, leading to lower significance. This suggests that the candidates with significance greater than $3\sigma$ may not exhibit periodic patterns in longer time series. The periods of these candidates range from 5 ms to 62 ms, reflecting the clustering of bursts within a 0.2-second time window. As expected, increasing the time window length makes it more difficult to maintain periodicity. This aligns with the results from searching for short-timescale periods in the TOA data, where pulse phases exhibited considerable spread rather than remaining confined to a narrow range. The 0.2-second clustering assumption implies that bursts occurring within such short intervals are likely to be physically connected, indicating the potential for periodicity in these local bursts.

\begin{table*}[htbp]
\footnotesize
\centering
\caption{QPO fitting parameters for individual bursts. The table lists the model components (Model), where components starting with G represent Gaussian models used to fit QPO peaks, and "Const" denotes the constant model representing the background level. The table also includes their associated amplitudes (Height), central frequencies (CF), full width at half maximum (FWHM), quality factors (Q), constant (C), and significance (Sig.) for the QPO in each individual burst. The MJD in each row indicates the start time of the time series for that particular burst.}
\label{tab:QPO_pars}
\begin{tabular}{c c c c c c c c c c}
\toprule
No. & MJD & DM & Sig. & Model & Height & CF & FWHM & Q & C ($\times 10^{-3}$) \\
    & From 60375 & $\rm pc/cm^3$ &  &  &  & Hz & Hz & & $\times 10^{-3}$ \\
\midrule
\multirow{4}{*}{1} & \multirow{4}{*}{0.182604131} & \multirow{4}{*}{528.5} & \multirow{4}{*}{3.4} 
& G$_0$ & 6.3370 $\pm$ 0.0460 & 4.45 $\pm$ 0.59 & 85.02 $\pm$ 0.63 & 0.0523 $\pm$ 0.0069 & - \\
&&&& G$_1$ & 0.1865 $\pm$ 0.0037 & 266.85 $\pm$ 0.68 & 97.00 $\pm$ 1.30 & 2.7500 $\pm$ 0.0370 & - \\
&&&& G$_2$ & 0.0053 $\pm$ 0.0011 & 502.10 $\pm$ 6.20 & 70.00 $\pm$ 14.00 & 7.1000 $\pm$ 1.4000 & - \\
&&&& Const & - & - & - & - & $3.0 \pm 4.6$ \\
\midrule
\multirow{4}{*}{2} & \multirow{4}{*}{1.143877879} & \multirow{4}{*}{528.7} & \multirow{4}{*}{3.7}
& G$_0$ & 36.3000 $\pm$ 0.4000 & 5.80 $\pm$ 1.10 & 118.50 $\pm$ 1.10 & 0.0487 $\pm$ 0.0094 & - \\
&&&& G$_1$ & 0.9460 $\pm$ 0.0210 & 327.94 $\pm$ 0.93 & 117.40 $\pm$ 1.70 & 2.7930 $\pm$ 0.0410 & - \\
&&&& G$_2$ & 0.0368 $\pm$ 0.0047 & 579.80 $\pm$ 5.60 & 105.00 $\pm$ 10.00 & 5.5000 $\pm$ 0.5500 & - \\
&&&& Const & - & - & - & - & $2.2 \pm 1.2$ \\
\bottomrule
\end{tabular}
\end{table*}

In the FFT analysis, after excluding radio frequency interference (RFI)—characterized by signals at DM = 0 or their narrowband nature—no periodic signal candidates were found. Nonetheless, two QPOs with significance greater than $3\sigma$ were detected. Figures \ref{fig:QPO1} and Figure \ref{fig:QPO2} display the two bursts containing QPOs. Each figure includes (a) the RFI-masked dynamic spectrum, (b) the dedispersed time series, (c) the periodogram with model fits under the null and alternative hypotheses, (d) residuals from the fits, (e) posterior distributions of model parameters from MCMC sampling, and (f) likelihood ratio test statistic distributions highlighting the observed values. In the periodograms, prominent excess power at specific frequencies indicates the presence of QPO components, with model residuals, MCMC posteriors, and likelihood ratio tests collectively confirming their statistical significance. To assess the statistical significance of the QPOs, we applied the Bayesian method described in Section \ref{sec:TS_method} and \ref{sec:appendix}. We performed hypothesis testing, where the null hypothesis $H_0$ assumed the periodogram could be fitted with one Gaussian model and one constant model, and the apparent QPOs were due to statistical fluctuations. The alternative hypothesis $H_1$ assumed that the statistical fluctuations under $H_0$ could not explain the apparent QPOs, requiring at least one additional Gaussian model for the QPOs. The LRT statistic yielded observed values of $T_{\rm LRT}^{\rm obs} = 34.2$ and $35.0$ for the two QPO events, respectively. Using MCMC sampling, we generated $1 \times 10^6$ simulated periodograms to compute the simulated likelihood ratios, which were compared with the observed value. As shown in Figure~\ref{fig:QPO1}f and Figure~\ref{fig:QPO2}f, the observed likelihood ratios for QPO1 and QPO2 lie in the tail of the simulated distributions under the null hypothesis, indicating that such values are unlikely to arise from statistical fluctuations alone. The corresponding $P$-values are $5.50 \times 10^{-4}$ and $1.53 \times 10^{-4}$, yielding detection significances of approximately $3.4\sigma$ and $3.7\sigma$, respectively. This result rejects the null hypothesis in favor of the alternative hypothesis, supporting the presence of significant QPOs. Table \ref{tab:QPO_pars} lists the fitting parameters of periodograms for the QPOs.

In summary, both the ACF and the FFT methods provided valuable insights into the periodic and quasi-periodic characteristics of the observed bursts. The ACF analysis revealed periodicity in short-timescale burst clustering, with characteristic timescales ranging from a few milliseconds to tens of milliseconds. In parallel, the FFT analysis identified QPOs with frequency components in the hundreds of Hz, thereby suggesting the presence of high-frequency oscillatory behavior. The complementary nature of these methods allows us to gain a comprehensive understanding of the periodic phenomena present in the burst time series.

\section{Discussion}\label{sec:discussion}
\subsection{Short-Timescale Periodicity in TOAs}\label{sec:short_timescale}
A natural expectation from magnetar-based models of repeating FRBs is the presence of underlying periodicities in burst arrival times that trace the rotation of the neutron star. Although a large-scale joint analysis revealed no significant global periodicity, consistent with the non-detection reported by the reference \cite{2022RAA....22l4004N}, we conducted detailed searches within each continuous observation and identified three sets of candidate periodicities, each observed in two separate epochs.  

Although we have identified three candidate periodic signals from FRB 20240114A, it is unlikely that all of them correspond to genuine periodicity to the source. Our current selection criteria do not permit us to definitively exclude any of the candidates. While the periods of these candidate signals are on the shorter side of the known magnetar period range, some known magnetars, such as PSR J1846-0258, have even shorter periods of $\sim$0.3266\,s \cite{2011ApJ...730...66L}.  

The observed magnitude of $\dot{P}$ can be used to estimate possible orbital configurations.
Assuming that the entire observed $\dot{f}/f \sim 10^{-7}-10^{-6}\ \mathrm{s^{-1}}$ is due to orbital motion, the corresponding orbital period for a companion mass of $M_c = 100 M_\odot$ would be approximately $P_{\rm orb} \sim 0.3-1.9$ days.
This range provides a rough constraint on the orbital parameters, indicating that if some of the rapid $\dot{P}$ variations are indeed caused by orbital effects, the orbital period would not be very long.

The absence of substantial periodic detections in our joint and clustered analyses suggests that any underlying periodicity is likely to be transient, intermittent, or strongly modulated, consistent with findings from other FRB studies \cite{2023A&A...678A.149P,2025arXiv250312013D}. Reference \cite{2025arXiv250312013D} proposes that FRB production may involve multiple independent emission regions in the magnetar magnetosphere, potentially explaining this intermittency. Should these periodicities be confirmed, they would suggest the presence of extremely young, highly magnetized neutron stars as FRB progenitors. If the binary system hypothesis holds, it could also shed light on the role of companion stars in modulating the magnetar's rotation. Future multi-wavelength observations will be crucial in verifying these periodicities and constraining the system's properties, including the potential presence of a companion star.

\subsection{Long-Timescale Periodicity in TOAs}
Our analysis of FRB 20240114A reveals a significant periodicity at $143.40 \pm 7.19$ days with $5.2\sigma$ significance. The coherent phase folding over the entire $\sim$20-month observational baseline demonstrates consistent burst concentration in specific phase regions, regardless of exposure correction. While minor differences exist between exposure-corrected and uncorrected profiles, both show prominent peaks at the same phases, supporting the astrophysical nature of this periodicity. The detection of a harmonically related component at approximately half the fundamental period further supports the astrophysical origin of this signal. The coherence of this periodicity is demonstrated by its consistent alignment with four consecutive active windows. However, the source entered a quiescent state during the final predicted active window. While our observations covered the beginning of this window up to MJD 60949—well within the $\sim$30-day active phase and just 8 days before the predicted peak at MJD 60957—no bursts were detected (see Figure \ref{fig:long_results}A). This demonstrates that while the periodic modulation is a robust and intrinsic property of the source, its expression as detectable bursting is contingent upon the state of a central engine that can transition into quiescence.

With this detection, FRB 20240114A joins the growing class of periodic repeating FRBs, which now exhibits a remarkable diversity in timescales. The population spans from the relatively short 16-day period of FRB 20180916B \cite{2020Natur.582..351C}, through the intermediate periods of FRB 20240209A (126 days; \cite{2025ApJ...983L..15P}) and the 143-day period reported here for FRB 20240114A, to the longer $\sim$160-day period of FRB 20121102A \cite{2020ApJ...893L..26I}. This diversity suggests that multiple physical mechanisms may be responsible for the observed periodicity across different sources.

Notably, the 143-day period of FRB 20240114A is remarkably close to the $\sim$160-day period of FRB 20121102A, suggesting these two highly active repeaters may share similar underlying physical mechanisms. Both sources exhibit complex burst morphologies and high activity levels, potentially indicating a common origin for their $\sim$150-day class periodicity. Meanwhile, the much shorter 16-day period of FRB 20180916B likely represents a distinct physical scenario.

For FRB 20240114A's 143-day period, binary orbital modulation provides a natural explanation. In such models, the period could correspond to orbital motion with either a massive stellar companion or a compact object \cite{2020MNRAS.498L...1Z,2021ApJ...918L...5L}. The similarity to FRB 20121102A's period suggests comparable binary separations or system configurations. To assess the consistency of this scenario with other timing features, we compared the expected orbital modulation effects with the properties of candidate short-timescale periodic signals identified in our analysis (Section~\ref{sec:short_timescale}). Adopting an extreme companion mass $M_c = 100 M_\odot$ for the 143-day orbital period produces a maximum fractional frequency derivative of $\left|\frac{\dot{f}}{f}\right|_{\rm max} \sim 10^{-10}\ \mathrm{s^{-1}}$. This value is several orders of magnitude smaller than the $\dot{f}/f$ variations ($\sim 10^{-7}-10^{-6}\ \mathrm{s^{-1}}$) measured for the candidate short-timescale signals in Table~\ref{tab:short_TDA}. This discrepancy indicates that if the 143-day period represents an orbital modulation, it operates on a distinct timescale from the processes responsible for the observed short-timescale timing variations. 

Alternatively, precession mechanisms—including free neutron star precession or jet precession in accreting systems—could produce modulation on such long timescales \cite{2020ApJ...892L..15Z,2021ApJ...921..147C}. The presence of harmonic structure in FRB 20240114A's phase profile may provide additional constraints on the system geometry.

Beyond activity periodicity, recent studies have revealed periodic evolution in rotation measures (RM), providing independent evidence for binary system origins. FRB 20201124A shows a 26.24-day RM periodicity \cite{2025arXiv250506006X}, while FRB 20220509 exhibits $\sim$200-day RM variations \cite{2025arXiv250510463L}. These RM periodicities strongly support binary system models where the FRB source moves through a magnetized plasma environment modulated by orbital motion.

The detection of this $\sim$143-day period establishes FRB 20240114A as a key member of the periodic FRB population. The existence of multiple period classes ($\sim$16 days, $\sim$126-160 days) suggests that periodicity may be a common feature among repeating FRBs, arising from different physical processes in diverse progenitor systems. The emerging picture from both activity and RM periodicities points toward binary system origins for many repeating FRBs, with orbital characteristics spanning a wide parameter space. Future multiwavelength observations and continued monitoring will be crucial for distinguishing between competing models and understanding the origin of these intriguing periodicities.

\subsection{Periodicity and QPOs in Burst Time Series}
By combining the ACF and the FFT analysis, we reveal temporal features in FRB emissions that are both distinct and complementary. 

Using the ACF-based method, we identified 4 candidates that exceed the $3\sigma$ threshold. Among them, only one shows four distinct pulses with clearly periodic burst times, indicating a regular temporal spacing. The detected periodicities (ranging from 5 to 62 ms) in the burst clusters exhibit an intriguing behavior: their statistical significance diminishes when the analysis is extended beyond 4 cycles. This suggests a form of transient coherence, possibly originating from unstable oscillation modes in the emission region. These modes may temporarily excite specific periodic components, leading to enhancements in periodicity. The lack of persistent periodic signals over longer timescales is consistent with the stochastic nature of burst arrival times. At the same time, the localized synchronization observed within short windows implies that some degree of phase correlation exists during active emission episodes. 

Using the FFT analysis, we identified 2 candidates exhibiting statistically significant QPO features. To further investigate the temporal characteristics of the two detected QPOs, we applied Empirical Mode Decomposition (EMD) to decompose their burst time series into multiple scales. EMD is an adaptive signal processing technique that decomposes complex nonlinear, non-stationary signals into a set of Intrinsic Mode Functions (IMFs), each representing a distinct time-frequency component \cite{1998RSPSA.454..903H,2020SMaS...29i3001B}. For the No. 1 QPO, the decomposition yielded five IMFs and one residual; for the No. 2 QPO, four IMFs and one residual were obtained. Based on the hierarchical structure of the IMFs from high to low frequency, we grouped IMFs 1–3 (for No. 1) or IMFs 1-2 (for No. 2) as Component 1 (white noise), IMFs 4–5 (for No. 1) or IMF 3-4 (for No. 2) as Component 2 (QPO component), and the residual as Component 3 (representing the low-frequency envelope of the burst). Periodograms were then calculated for each component.

Figures~\ref{fig:EMD_QPO}a–d in \ref{appendix:QPO} show the original burst time series and the three components for the No. 1 QPO, while Figures~\ref{fig:EMD_QPO}e–h display their corresponding periodograms. The same analysis for the No. 2 QPO is shown in Figures~\ref{fig:EMD_QPO}i–l and \ref{fig:EMD_QPO}m–p. Component 1 mainly contains white noise, Component 3 reflects the overall envelope of the burst profile, and Component 2 exhibits clear oscillatory behavior. The periodogram of Component 2 matches the original QPO periodogram, confirming that this component captures the QPO signal extracted through EMD.

To characterize the time-domain behavior of the QPOs, we adopted the fitting method used by the reference \cite{2022ApJ...931...56L}, in which the QPO is modeled as the product of a Gaussian envelope and a sinusoidal oscillation. This approach assumes that QPOs are amplitude-modulated signals, described by the following expression:
\begin{equation}
I(t) = A e^{-\frac{(t - t_c)^2}{2\sigma^2}} \sin(2\pi f t + \phi),
\label{eq:fit_QPO}
\end{equation}
where $A$ is the peak amplitude, $t_c$ is the center of the Gaussian function, $\sigma$ is its width, $f$ is the oscillation frequency, and $\phi$ is the phase. Using this model to fit the data, we obtain the optimal parameters for the No. 1 QPO as
$A = 1.050$, $t_c = 0.0260\, \mathrm{s}$, $\sigma = 0.0028\, \mathrm{s}$, $f = 270.40\, \mathrm{Hz}$, and $\phi = -2.30\, \mathrm{rad}$. For the No. 2 QPO, the fit yields
$A = 0.534$, $t_c = 0.0225\, \mathrm{s}$, $\sigma = 0.00235\, \mathrm{s}$, $f = 324.42\, \mathrm{Hz}$, and $\phi = -3.14\, \mathrm{rad}$. In both cases, the fitted oscillation frequency $f$ is consistent with the peak frequencies obtained from the corresponding periodograms, further confirming the presence of QPOs in the original burst time series.

The microstructure of FRBs provides crucial clues for understanding their radiation mechanisms and the physical properties of their central engines. Among these features, the discovery of QPOs is particularly intriguing. The presence of QPOs may be closely related to various physical processes in compact objects, such as rotation, gravitational lensing, binary star mergers, crustal oscillations in magnetars, or instabilities in the magnetosphere \cite{2012ApJ...755...80P,2016ApJ...822L...7W,2020ApJ...903L..38W,2023A&A...678A.149P}. These periodic or quasi-periodic phenomena likely reflect the dynamic behavior of compact objects under extreme magnetic fields or the oscillation modes of magnetospheric plasma. In-depth studies of these QPO features hold the potential to offer new insights into the origin of FRB radiation and their connection to compact objects.

A possible physical scenario involves a periodic pulse train superimposed on a broader pulse envelope, which could arise from temporal modulation of the signal strength (e.g., through gravitational lensing effects in pulsars). This mechanism naturally produces a bell-shaped overall profile with short-timescale periodic sub-structures, and may also be considered in the context of FRB emission models. In contrast, the pulse signals we observed do not exhibit strict periodicity, instead showing QPOs. Therefore, the gravitational lensing model is even less effective in explaining the phenomenon we observed. They also explored the possibility of explaining the periodic signal through neutron star mergers. Neutron star mergers are expected to produce QPOs ranging from a few Hz to several kHz, although the merger process is extremely brief. The duration typically depends on the companion star's mass, ranging from approximately $\sim10^2-10^4$ s, or between 0.1 and 1 s \cite{2022Natur.607..256C}. In contrast, FRB 20240114A is a repeater, which has been erupting for over a year, with no signs suggesting that the bursts will end in the short term. Therefore, the QPOs of FRB 20240114A cannot originate from a neutron star merger.

The microstructure of FRB pulses, especially the appearance of QPOs, may be related to the magnetar's X-ray burst activity and could be closely associated with the magnetar's vibrations or intense magnetic field activity \cite{2017ApJ...841...54T, 2019MNRAS.488.5887S, 2021NatAs...5..378L, 2022ApJ...931...56L}. In this context, recent three-dimensional simulations of magnetar crustal quakes \cite{2025arXiv250812567Q} suggest a plausible origin for the observed QPOs. The $\sim$ kHz seismic oscillations are found to bounce within the crust on millisecond timescales, which may account for the hundreds-of-hertz QPOs detected in our observations and in other FRBs. Furthermore, their simulation demonstrates that such a quake launches a mixture of Alfv\'en and fast magnetosonic waves into the magnetosphere. The damping of crustal motions on millisecond timescales due to energy loss to the core aligns with the ``transient coherence'' and short-lived nature of the periodicities we observed. Since the seismic waves are damped before they can spread laterally, the resulting magnetospheric emission remains confined near the quake’s epicenter. This localization may explain why high-significance QPOs appear only intermittently in specific bursts.

The timescale of the microstructure in the radio pulses of pulsar radiation is closely related to the pulsar's rotation period. Previous studies have shown that for normal, non-recycled pulsars, the quasi-period in their radio pulses is approximately three orders of magnitude smaller than their spin period, i.e., $P_\mu\simeq 10^{-3}P$, where $P_\mu$ and $P$ are the quasi-period and the spin period of the pulsar's radio pulses, respectively. Since we detected some QPOs in FRB 20240114A, if it indeed originates from a magnetar, we can attempt to estimate the magnetar's period based on the known relationship between the quasi-period of substructures and the spin period. The relationship is given by $P_{\mu}=(0.94\pm 0.04)\times P^{0.97\pm0.05} \, \text{ms}$, which has been confirmed in both normal pulsars and magnetars and is applicable over about six orders of magnitude \cite{2024NatAs...8..230K}. Following the method in the reference \cite{2024NatAs...8..230K}, we computed the geometric mean of all detected quasi-periods and estimated the magnetar's spin period to be in the range of approximately 3 s to 76 s.

The detection of QPOs in only a handful of bursts among tens of thousands cannot be easily explained by sample size or signal-to-noise considerations alone. This sporadic appearance aligns with previous observations that QPOs are not always present, even under similar conditions. For instance, the magnetar PSR J1622–4950 displayed clear QPO-like structures in 2017 observations, yet they were much weaker or absent in data from early and late 2018 \cite{2024NatAs...8..230K}, reinforcing the notion that their occurrence is inherently variable rather than purely observational in origin.

\section{Summary}\label{sec:summary}
This study presents a multi-timescale analysis of (quasi-)periodic signals in the repeating fast radio burst FRB~20240114A. The key findings are:
\begin{enumerate}
    \item Based on TOA data from 57 FAST observation sessions spanning UT 2024-01-28 to UT 2024-08-29, we identified three candidate short-timescale periodic signals with periods of 0.673 s (1.486 Hz), 0.635 s, and 0.536 s, with significances of $3.2$–$6\sigma$, each detected in two independent observations. It is clear that not all of these signals are related to the intrinsic periodicity of the source. Nevertheless, our current selection criteria are not sufficient to definitively rule out any of them.

    \item Using FAST observations with TOAs spanning from UT 2024-01-28 to UT 2025-10-01, we detected a significant long-timescale periodicity at $143.40 \pm 7.19$ days with $5.2\sigma$ significance. The coherent phase folding over the entire $\sim$20-month observational baseline demonstrates consistent burst concentration in specific phase regions, establishing FRB~20240114A as a key member of the periodic repeater population.

    \item Based on individual burst time series data from the 57 FAST observations spanning from UT 2024-01-28 to UT 2024-08-29, we identified 2 QPOs with significance of 3.4$\sigma$ and 3.7$\sigma$, respectively, in the few hundred Hz range. Using burst cluster time series data, autocorrelation analysis revealed 4 periodic candidates with periods ranging from a few milliseconds to tens of milliseconds, each with significance between 3$\sigma$ and 3.9$\sigma$. The periodicity quickly diminished as more bursts were added, suggesting that the periodicity is very short-lived.
\end{enumerate}

FRB~20240114A is currently one of the most active known repeating fast radio bursts, and our periodicity analysis provides important statistical constraints on its physical properties. Interestingly, FRB~20240114A exhibits a fascinating self-similar hierarchical structure in its timing properties: on the longest timescale, we observe a $\sim$143-day periodicity in activity windows; within individual active episodes, bursts themselves show short-timescale periodicities of $\sim$tens of milliseconds; and finally, at the finest scale, individual bursts display quasi-periodic oscillations on timescales of a few milliseconds. The current observational dataset is the largest for a single FRB to date, yet the lack of a detected spin period may be attributed to the intrinsic properties of the source rather than data limitations. Our findings provide valuable insights into the physical mechanisms of FRBs and will contribute to clarifying the potential connection among the observed periodicity and the source's rotation, orbital motion, and precession.
\Acknowledgements{This work made use of the data from FAST FRB Key Science Project. This work is supported by the National Natural Science Foundation of China (NSFC) under grant numbers 12588202, 12303042, 12041303, 12421003, 12233002, 12041306, 12403100, W2442001, 12203045, 12447115, U2031117 and the National SKA Program of China (2020SKA0120100, 2020SKA0120200, 2022SKA0130104). Pei Wang acknowledges support from the Youth Innovation Promotion Association CAS (id. 2021055), CAS Youth Interdisciplinary Team. Di Li is supported by the International Partnership Program of Chinese Academy of Sciences, Program No.114A11KYSB20210010, National Key R\&D Program of China No. 2023YFE0110500, QN2023061004L. Yi Feng is supported by the Leading Innovation and Entrepreneurship Team of Zhejiang Province of China grant No. 2023R01008, and by Key R\&D Program of Zhejiang grant No. 2024SSYS0012. Yongfeng Huang also acknowledges the support from the Xinjiang Tianchi Program. Qin Wu is supported by the China Postdoctoral Science Foundation (CPSF) under grant numbers GZB20240308, 2025T180875. C.W.T is supported by CAS project No. JZHKYPT-2021-06. J. R. Niu is supported by the National Natural Science Foundation of China (NSFC, No. 12503055) and the Postdoctoral Fellowship Program of CPSF under Grant Number GZB20250737. This research is also supported by the CAS Project for Young Scientists in Basic Research, YSBR-063.}

\InterestConflict{The authors declare that they have no conflict of interest.}


\bibliographystyle{scpma} 
\bibliography{ref}
\end{multicols}
\clearpage

\begin{appendix}
\section{Performance of the TDA Pipeline on Simulated TOA Data}
\label{appendix:TDA}
To evaluate the Time-Differencing Algorithm (TDA) in short-timescale periodicity searches, we simulated 300 TOAs over 1500~s with injected parameters $f_0 = 3.1415~\mathrm{Hz}$ and $f_1 = 1.0\times10^{-6}~\mathrm{Hz/s}$, resembling the observational characteristics of FRB\,20240114A. The analysis followed the main text pipeline: a TDA pre-screening over $f_1/f_0$ to locate candidate periodicities, followed by refinement using the $Z_n^2$ statistic.
\begin{figure}[h!]
    \centering
    \includegraphics[scale=0.43]{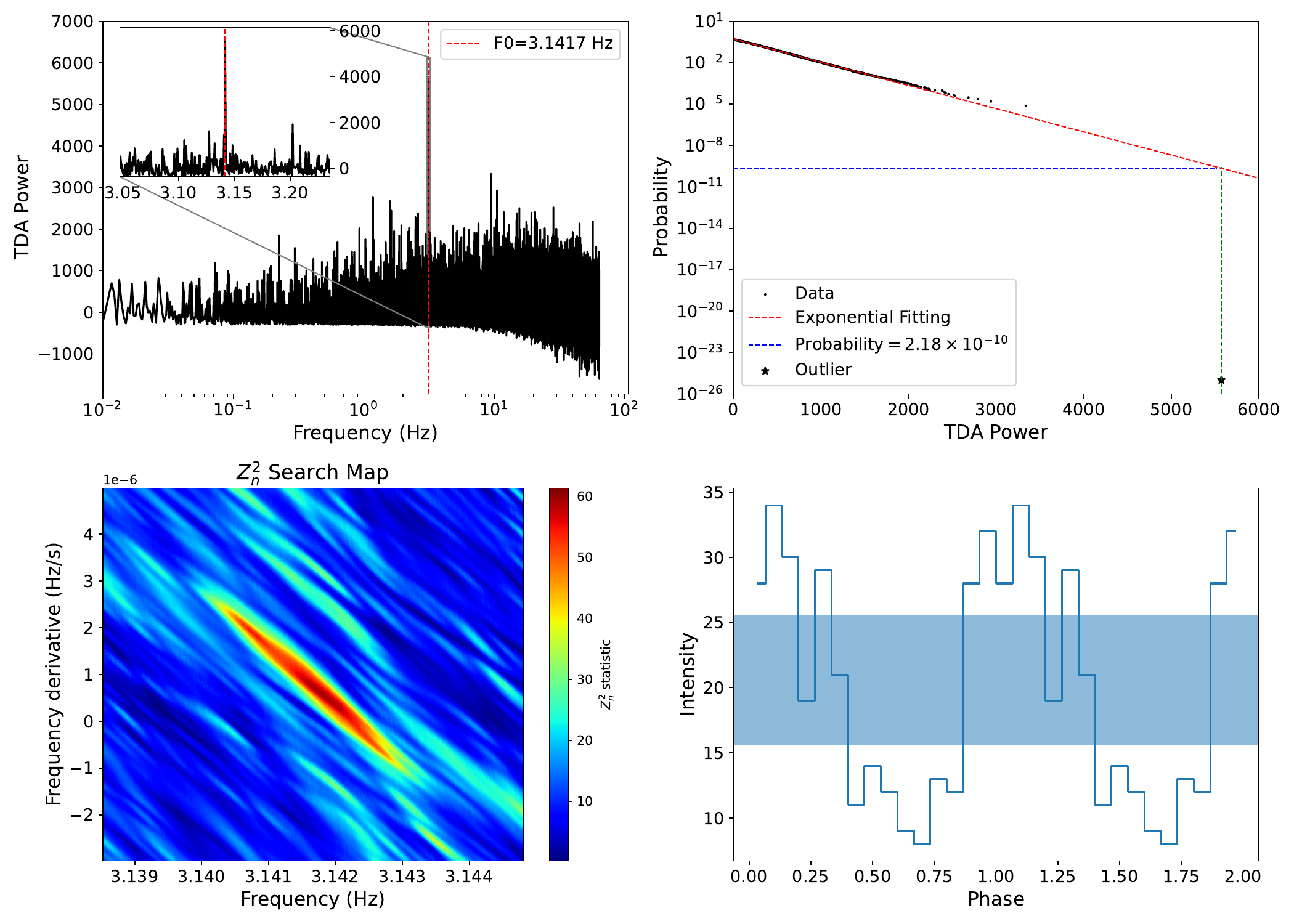}
    \caption{TDA pre-screening followed by $Z_n^2$ refinement on simulated TOA data. 
    \textbf{Top left}: TDA spectrum showing the outlier near $f_0 = 3.1417$~Hz corresponding to the peak at $f_1/f_0 = 3.1679\times10^{-7}\rm s^{-1}$. 
    \textbf{Top right}: Exponential fit to the CCDF of the TDA spectrum, yielding $6.34\sigma$ significance. 
    \textbf{Bottom left}: $Z_n^2$ map using TDA-initialized parameters, highlighting the maximum near the true values. 
    \textbf{Bottom right}: Phase-folded TOAs using $Z_n^2$ results, showing clear phase clustering and blue bars indicating the 1$\sigma$ range expected from pure Poisson noise.}
    \label{fig:tda_z2}
\end{figure}

Using this pipeline, the most significant TDA peak was found at $f_1/f_0 = 3.1679\times10^{-7}\rm s^{-1}$, corresponding to an outlier near $f_0 = 3.1417~\mathrm{Hz}$ (Figure~\ref{fig:tda_z2}, top left). From this $f_0$, the estimated frequency derivative is $f_1 = 9.9525\times10^{-7}~\mathrm{Hz/s}$, closely matching the injected values. Fitting the complementary cumulative distribution function (CCDF) of the TDA spectrum with an exponential model (top right) yields a false-alarm probability of $2.18\times10^{-10}$, corresponding to a $6.34\sigma$ significance.

Refinement via $Z_n^2$ (bottom left) produces a localized maximum at $f_0 = 3.1414~\mathrm{Hz}$ and $f_1 = 1.007\times10^{-6}~\mathrm{Hz/s}$, providing an accurate recovery of the injected parameters. Phase-folding the TOAs using these refined values (bottom right) shows clear clustering within a narrow phase window, with blue bars indicating the 1$\sigma$ range expected from pure Poisson noise.

These simulations demonstrate that TDA effectively identifies candidate periodicities in sparse TOA datasets. When combined with $Z_n^2$ refinement, the hybrid approach achieves both computational efficiency and accurate recovery of injected periods, supporting its application for short-timescale periodicity searches in FRB-like datasets.

\section{FAST Observational Log of FRB 20240114A}
\label{app:observing_log}

\begin{center}
\scriptsize
\begin{longtable}{lrrrrr}
\caption{FAST observational log of FRB 20240114A used in the long-timescale periodicity analysis. The table lists observation dates, MJD for start and end times, observation durations in hours, number of detected bursts, and burst rates in bursts per hour.}
\label{tab:observing_log} \\
\hline
Date (UT) & Start MJD & End MJD & Duration (hr) & Burst Count & Rate (hr$^{-1}$) \\
\hline
\endfirsthead

\caption[]{Continued.} \\
\hline
Date & Start MJD & End MJD & Duration (hr) & Burst Count & Rate (hr$^{-1}$) \\
\hline
\endhead

\hline
\multicolumn{6}{r}{Continued on next page} \\
\endfoot

\hline
\endlastfoot
20240128 & 60337.229861 & 60337.250694 & 0.50 & 8 & 16.0 \\
20240129 & 60338.215278 & 60338.236111 & 0.50 & 11 & 22.0 \\
20240201 & 60341.179861 & 60341.200694 & 0.50 & 9 & 18.0 \\
20240204 & 60344.240972 & 60344.261806 & 0.50 & 2 & 4.0 \\
20240209 & 60349.152778 & 60349.173611 & 0.50 & 6 & 12.0 \\
20240213 & 60353.215972 & 60353.236806 & 0.50 & 14 & 28.0 \\
20240215 & 60355.131250 & 60355.152083 & 0.50 & 21 & 42.0 \\
20240221 & 60361.179167 & 60361.200000 & 0.50 & 24 & 48.0 \\
20240225 & 60365.131944 & 60365.152778 & 0.50 & 11 & 22.0 \\
20240305 & 60374.188194 & 60374.209028 & 0.50 & 225 & 450.0 \\
20240306 & 60375.134722 & 60375.218056 & 2.00 & 918 & 459.0 \\
20240307 & 60376.102778 & 60376.186111 & 2.00 & 969 & 484.5 \\
20240308 & 60377.193056 & 60377.221458 & 0.68 & 347 & 509.0 \\
20240309 & 60378.191667 & 60378.205671 & 0.34 & 186 & 553.4 \\
20240310 & 60379.199306 & 60379.219132 & 0.48 & 270 & 567.4 \\
20240311 & 60380.152778 & 60380.194444 & 1.00 & 727 & 727.0 \\
20240312 & 60381.038194 & 60381.220833 & 4.38 & 3197 & 729.4 \\
20240313 & 60382.173611 & 60382.215278 & 1.00 & 535 & 535.0 \\
20240315 & 60384.190972 & 60384.204861 & 0.33 & 66 & 198.0 \\
20240317 & 60386.118750 & 60386.139583 & 0.50 & 171 & 342.0 \\
20240321 & 60390.158333 & 60390.172222 & 0.33 & 111 & 333.0 \\
20240327 & 60396.133333 & 60396.154167 & 0.50 & 297 & 594.0 \\
20240329 & 60398.981944 & 60399.002778 & 0.50 & 275 & 550.0 \\
20240401 & 60401.066667 & 60401.087500 & 0.50 & 261 & 522.0 \\
20240403 & 60403.046528 & 60403.067361 & 0.50 & 141 & 282.0 \\
20240407 & 60407.091667 & 60407.112500 & 0.50 & 41 & 82.0 \\
20240410 & 60410.020833 & 60410.041667 & 0.50 & 35 & 70.0 \\
20240413 & 60413.023611 & 60413.044444 & 0.50 & 44 & 88.0 \\
20240417 & 60417.078472 & 60417.099306 & 0.50 & 45 & 90.0 \\
20240421 & 60421.009028 & 60421.029861 & 0.50 & 72 & 144.0 \\
20240424 & 60424.983333 & 60425.004167 & 0.50 & 110 & 220.0 \\
20240430 & 60430.029167 & 60430.050000 & 0.50 & 133 & 266.0 \\
20240504 & 60434.956944 & 60434.977778 & 0.50 & 89 & 178.0 \\
20240511 & 60441.888889 & 60441.909722 & 0.50 & 102 & 204.0 \\
20240518 & 60448.897222 & 60448.911111 & 0.33 & 32 & 96.0 \\
20240525 & 60455.913889 & 60455.927778 & 0.33 & 37 & 111.0 \\
20240531 & 60461.949306 & 60461.963194 & 0.33 & 11 & 33.0 \\
20240608 & 60469.827778 & 60469.841667 & 0.33 & 49 & 147.0 \\
20240615 & 60476.877083 & 60476.890972 & 0.33 & 29 & 87.0 \\
20240622 & 60483.881944 & 60483.895833 & 0.33 & 40 & 120.0 \\
20240629 & 60490.862500 & 60490.876389 & 0.33 & 46 & 138.0 \\
20240706 & 60497.858333 & 60497.872222 & 0.33 & 42 & 126.0 \\
20240713 & 60504.761111 & 60504.775000 & 0.33 & 117 & 351.0 \\
20240720 & 60511.684722 & 60511.698611 & 0.33 & 100 & 300.0 \\
20240722 & 60513.800694 & 60513.815278 & 0.35 & 92 & 262.9 \\
20240724 & 60515.773611 & 60515.834722 & 1.47 & 369 & 251.6 \\
20240726 & 60517.819444 & 60517.840278 & 0.50 & 166 & 332.0 \\
20240727 & 60518.742361 & 60518.756250 & 0.33 & 92 & 276.0 \\
20240728 & 60519.765278 & 60519.786111 & 0.50 & 208 & 416.0 \\
20240730 & 60521.800000 & 60521.820833 & 0.50 & 138 & 276.0 \\
20240802 & 60524.783333 & 60524.797222 & 0.33 & 108 & 324.0 \\
20240805 & 60527.777778 & 60527.791667 & 0.33 & 52 & 156.0 \\
20240808 & 60530.775000 & 60530.788889 & 0.33 & 90 & 270.0 \\
20240814 & 60536.777778 & 60536.791667 & 0.33 & 86 & 258.0 \\
20240817 & 60539.709722 & 60539.723611 & 0.33 & 70 & 210.0 \\
20240821 & 60543.737500 & 60543.751389 & 0.33 & 66 & 198.0 \\
20240829 & 60551.699306 & 60551.713194 & 0.33 & 40 & 120.0 \\
20240905 & 60558.693750 & 60558.707639 & 0.33 & 54 & 162.0 \\
20240913 & 60566.535417 & 60566.549306 & 0.33 & 78 & 234.0 \\
20240919 & 60572.679167 & 60572.693056 & 0.33 & 33 & 99.0 \\
20240927 & 60580.632639 & 60580.646528 & 0.33 & 37 & 111.0 \\
20241005 & 60588.527778 & 60588.541667 & 0.33 & 34 & 102.0 \\
20241011 & 60594.479884 & 60594.493773 & 0.33 & 38 & 114.0 \\
20241021 & 60604.520833 & 60604.534722 & 0.33 & 51 & 153.0 \\
20241029 & 60612.475000 & 60612.488889 & 0.33 & 41 & 123.0 \\
20241105 & 60619.415972 & 60619.429861 & 0.33 & 26 & 78.0 \\
20241116 & 60630.405556 & 60630.488889 & 2.00 & 283 & 141.5 \\
20241123 & 60637.473611 & 60637.487500 & 0.33 & 47 & 141.0 \\
20241204 & 60648.402778 & 60648.416667 & 0.33 & 43 & 129.0 \\
20241209 & 60653.353472 & 60653.436806 & 2.00 & 332 & 166.0 \\
20241216 & 60660.434028 & 60660.447917 & 0.33 & 31 & 93.0 \\
20241225 & 60669.347917 & 60669.431250 & 2.00 & 521 & 260.5 \\
20241226 & 60670.399306 & 60670.413194 & 0.33 & 115 & 345.0 \\
20250101 & 60676.313889 & 60676.381944 & 1.63 & 653 & 399.8 \\
20250102 & 60677.269444 & 60677.283333 & 0.33 & 113 & 339.0 \\
20250108 & 60683.306250 & 60683.347917 & 1.00 & 186 & 186.0 \\
20250112 & 60687.291667 & 60687.333333 & 1.00 & 475 & 475.0 \\
20250115 & 60690.245833 & 60690.329167 & 2.00 & 958 & 479.0 \\
20250119 & 60694.281250 & 60694.302083 & 0.50 & 210 & 420.0 \\
20250123 & 60698.219444 & 60698.233333 & 0.33 & 79 & 237.0 \\
20250127 & 60702.218750 & 60702.232639 & 0.33 & 35 & 105.0 \\
20250131 & 60706.240972 & 60706.282639 & 1.00 & 82 & 82.0 \\
20250208 & 60714.268056 & 60714.281944 & 0.33 & 17 & 51.0 \\
20250215 & 60721.217361 & 60721.259028 & 1.00 & 118 & 118.0 \\
20250222 & 60728.248611 & 60728.262500 & 0.33 & 23 & 69.0 \\
20250301 & 60735.145139 & 60735.159028 & 0.33 & 20 & 60.0 \\
20250308 & 60742.118750 & 60742.132639 & 0.33 & 46 & 138.0 \\
20250317 & 60751.089583 & 60751.103472 & 0.33 & 40 & 120.0 \\
20250322 & 60756.065278 & 60756.079167 & 0.33 & 39 & 117.0 \\
20250328 & 60762.997222 & 60763.011111 & 0.33 & 33 & 99.0 \\
20250423 & 60788.921528 & 60788.935417 & 0.33 & 8 & 24.0 \\
20250501 & 60796.894444 & 60796.908333 & 0.33 & 57 & 171.0 \\
20250508 & 60803.979861 & 60803.993750 & 0.33 & 156 & 468.0 \\
20250522 & 60817.883333 & 60817.897222 & 0.33 & 51 & 153.0 \\
20250525 & 60820.976389 & 60820.990278 & 0.33 & 49 & 147.0 \\
20250529 & 60824.882639 & 60824.896528 & 0.33 & 46 & 138.0 \\
20250605 & 60831.968056 & 60831.981944 & 0.33 & 1 & 3.0 \\
20250612 & 60838.804167 & 60838.818056 & 0.33 & 0 & 0.0 \\
20250621 & 60847.756944 & 60847.798611 & 1.00 & 0 & 0.0 \\
20250624 & 60850.750000 & 60850.770833 & 0.50 & 0 & 0.0 \\
20250713 & 60869.746528 & 60869.760417 & 0.33 & 0 & 0.0 \\
20250720 & 60876.792361 & 60876.834028 & 1.00 & 0 & 0.0 \\
20250721 & 60877.711111 & 60877.731944 & 0.50 & 0 & 0.0 \\
20250723 & 60879.781250 & 60879.802083 & 0.50 & 0 & 0.0 \\
20250725 & 60881.723611 & 60881.744444 & 0.50 & 0 & 0.0 \\
20250813 & 60900.753472 & 60900.767361 & 0.33 & 0 & 0.0 \\
20250824 & 60911.720139 & 60911.734028 & 0.33 & 0 & 0.0 \\
20250901 & 60919.604861 & 60919.618750 & 0.33 & 0 & 0.0 \\
20250912 & 60930.589583 & 60930.603472 & 0.33 & 0 & 0.0 \\
20250921 & 60939.646528 & 60939.660417 & 0.33 & 0 & 0.0 \\
20251001 & 60949.559722 & 60949.573611 & 0.33 & 0 & 0.0 \\
\end{longtable}
\end{center}
\normalsize

\section{Search Method for Periodic Signals and QPOs in bursts}\label{sec:appendix}
To reliably identify periodic signals and quasi-periodic oscillations (QPOs) in the bursts, we adopt the method proposed by \cite{2013ApJ...768...87H}, which is based on Bayesian inference. The procedure begins by computing the periodogram of the burst time series using the Fast Fourier Transform (FFT). A simpler model, referred to as the null hypothesis model, is then selected, and the periodogram is fit using maximum likelihood estimation (MLE) to obtain the a maximum posteriori (MAP) parameters under this null hypothesis. The analysis is then performed separately for searching periodic signals and QPOs.

For searching periodic signals, the residual at each frequency is computed as:
\begin{equation}
    R_j = \frac{2I_j}{S_j},
\end{equation}
where \(I_j\) and \(S_j\) are the values of the periodogram and the null hypothesis model at frequency \(f_j\), respectively. The maximum residual, \(\max(R_j)\), is identified as a potential periodic signal. The likelihood under the null hypothesis model is calculated as:
\begin{equation}
        p\left(\bm{I}|\hat{\theta}_{\rm MAP}^0, H_0\right) = \prod\limits_{i=1}^{n} p\left(I_j | S_j\right),
\end{equation}
where \(p(I_j | S_j)\) represents the probability density function (PDF) of the periodogram at frequency \(f_j\) given the model \(S_j\). Given the model $S_j$, the probability density of the power $I_j$ is \cite{van1989fourier,1995A&A...300..707T,2010MNRAS.402..307V}
\begin{equation}
    p(I_j|S_j)=\frac{1}{S_j}e^{-\frac{I_j}{S_j}}.
\end{equation}
The posterior distribution of the null hypothesis model parameters is computed by combining this likelihood with the prior distribution. MCMC sampling is used to obtain a large number of simulated periodograms based on the posterior distribution. The simulated periodograms are then fit with the null hypothesis model, and the maximum residuals for each simulated periodogram are calculated. The \(P\)-value is determined by comparing the maximum residual observed in the actual burst periodogram with those from the simulated ones. If the \(P\)-value is below a predefined significance level, the outlier is considered statistically significant and warrants further investigation.

For searching QPOs, the periodogram is fitted with a more complex nested model, referred to as the alternative hypothesis model, to obtain the MAP parameters under this hypothesis. The likelihood ratio is computed by comparing the likelihoods of the periodogram given both the null and alternative hypothesis models:
\begin{equation}
        T_{\rm LRT}^{\rm obs} = -2\log \frac{p\left(\bm{I}|\hat{\theta}_{\rm MAP}^0, H_0\right)}{p\left(\bm{I}|\hat{\theta}_{\rm MAP}^1, H_1\right)}.
\end{equation}
MCMC sampling is also used here to generate simulated periodograms based on the null hypothesis model. The likelihood ratios are calculated for each simulated periodogram, and the \(P\)-value is determined by comparing the observed likelihood ratio with the sample distribution obtained from the simulations.

Our search strategy follows that of \cite{2013ApJ...768...87H} with the following key adjustments: the null hypothesis model consists of a Gaussian model and a constant function model. The Gaussian model fits the low-frequency component of the periodogram, while the constant function model fits the baseline of the periodogram. We found that the Gaussian model provides the best fit for the low-frequency component, and additional Gaussian models are added to fit narrower features in the periodogram. This results in a nested model that serves as the alternative hypothesis. After the initial candidate is identified, the periodogram is reviewed manually to determine if additional Gaussian models are needed. If the periodogram cannot be fit with at most five Gaussian functions plus one constant function (with a reduced \(\chi^2 > 1.5\)), the candidate is discarded.

While the $P$-values reported in this work already account for multiple testing across frequencies in each individual periodogram, we do not apply additional corrections for multiple bursts since we focus on assessing significance within each burst individually. This conservative approach avoids over-correction, especially since previously published bursts are not fully included in our analysis.

\clearpage
\section{Supplementary Figures for Quasi-Periodic Signal Analysis}\label{appendix:QPO}
\begin{figure*}[h!]
\centering
\includegraphics[scale=0.60]{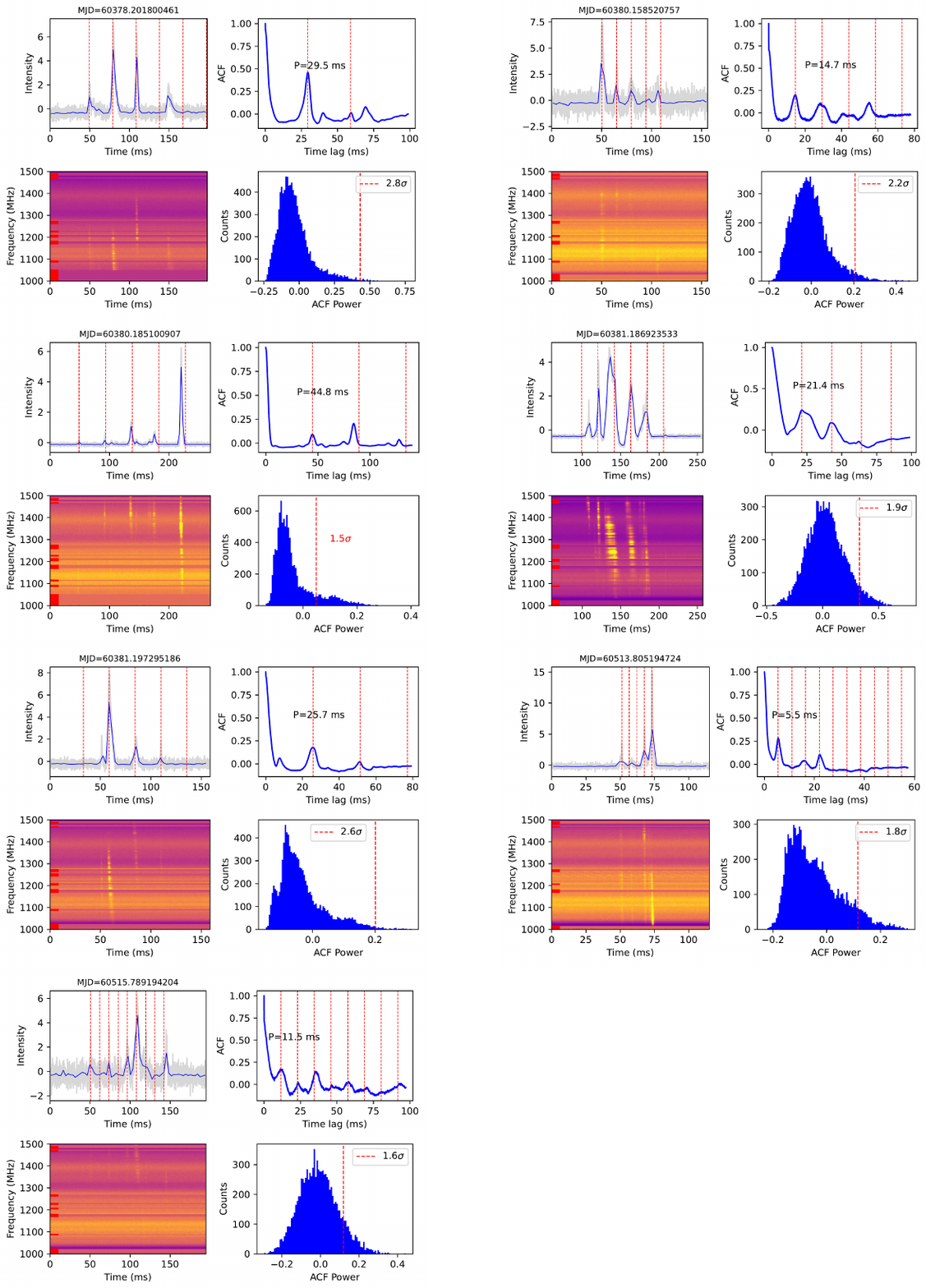}
\caption{Candidate periodic signals identified from the burst time series using the ACF method, each showing periodic indications with some significance, although not exceeding $3\sigma$. Each candidate consists of four subplots: the top-left subplot shows the burst time series, the bottom-left subplot shows the corresponding dynamic spectrum, the top-right subplot shows the ACF of the burst time series, and the bottom-right subplot shows the simulated statistic used to calculate the probability of generating this ACF pattern. The red line in the burst time series represents the periodic intervals determined by the ACF analysis, while the red line in the simulated statistic corresponds to the observed value.}
\label{fig:ACF_2}
\end{figure*}

\begin{figure*}
    \centering
    \includegraphics[scale=0.7]{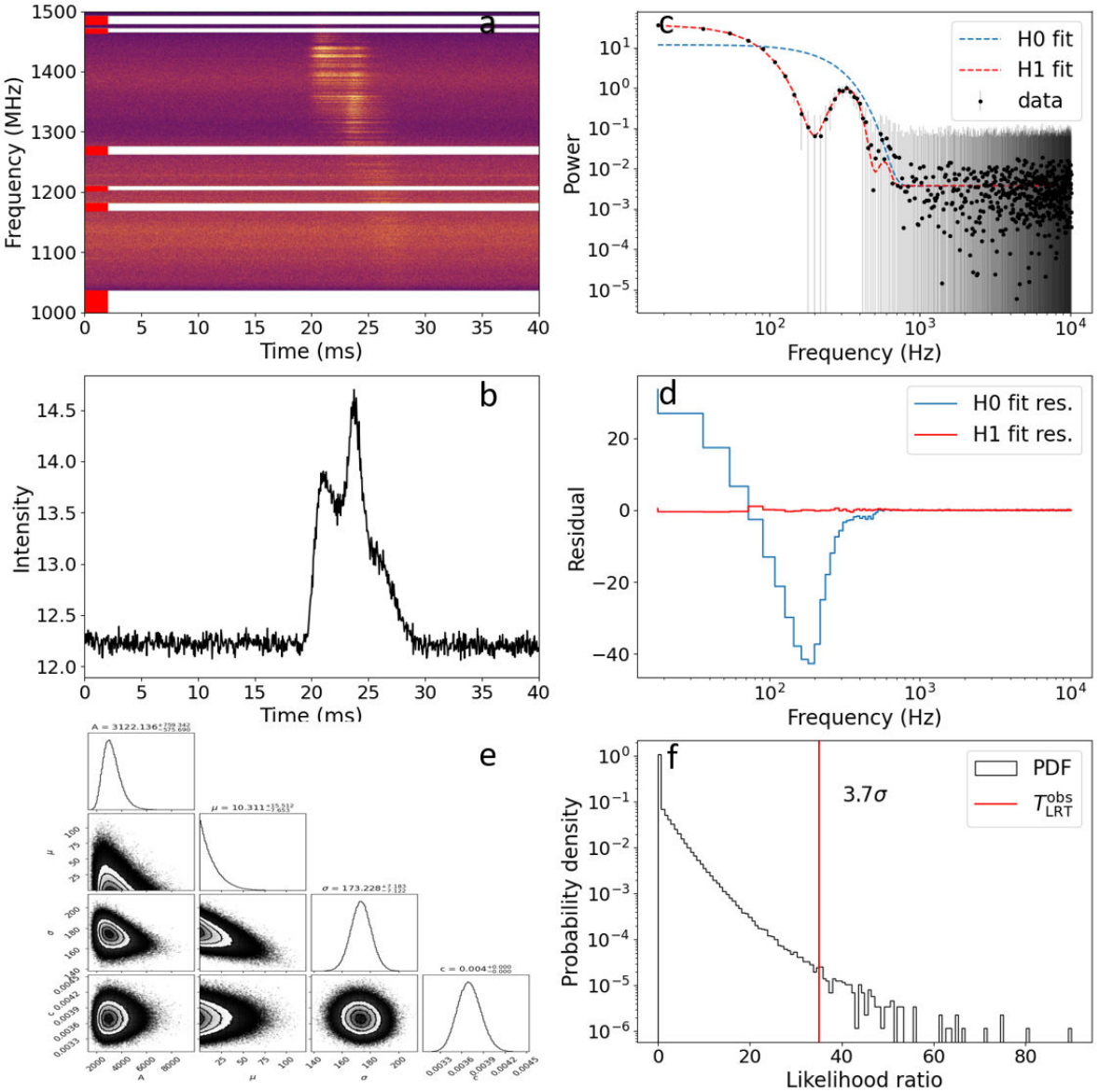}
    \caption{Same as Figure~\ref{fig:QPO1}, but for the burst corresponding to entry No.\,2 in Table~\ref{tab:QPO_pars}.}
    \label{fig:QPO2}
\end{figure*}

\begin{figure*}
    \centering
    \includegraphics[scale=0.47]{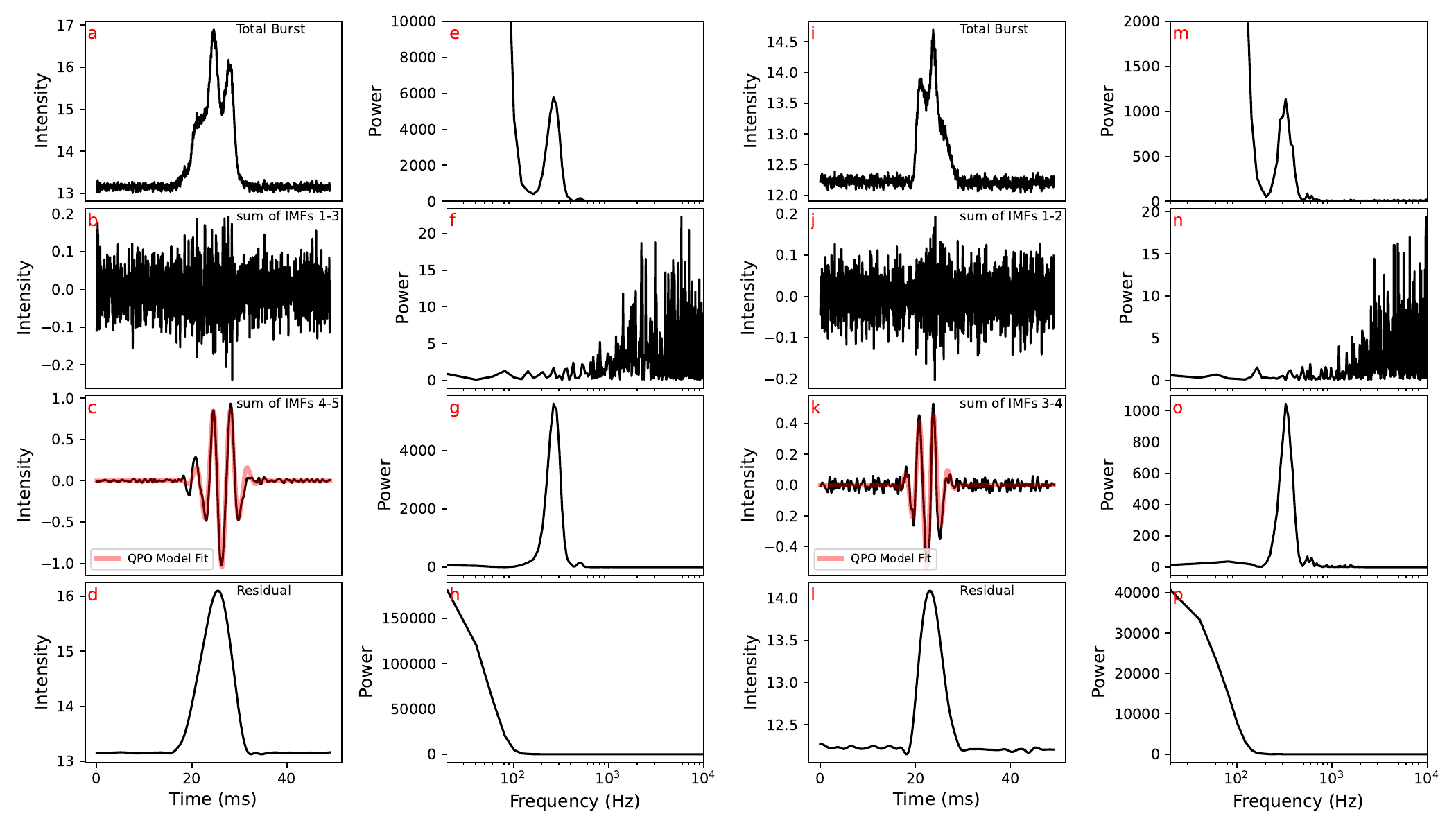}
    \caption{EMD results for the two detected QPO burst time series listed in Table \ref{tab:QPO_pars} (No. 1 and No. 2 QPOs). Panels a–d show the results for the No. 1 QPO: (a) the original burst time series, (b) the white noise component reconstructed by summing IMFs 1–3, (c) the QPO signal component from IMFs 4–5, and (d) the residual representing the burst envelope. Panels e–h show the corresponding periodograms of the time series in panels a–d. Panels i–l and m–p present the same analysis for the No. 2 QPO. The pink curves in panels c and k show the best-fit QPO time-domain signals using model (\ref{eq:fit_QPO}).}
    \label{fig:EMD_QPO}
\end{figure*}

\end{appendix}

\end{document}